# Fast reduction of electron-beam-activated graphene oxide by an infrared laser pulse

Israt Ali [a], Hilaire Mba [b], Matthieu Picher [b], Shruti Verma [a], Florian Banhart [b], and Kenneth R. Beyerlein [a]*

[a] Institut National de la Recherche Scientifique (INRS), Center Énergie Matériaux Télécommunications, Varennes, Québec, J3X 1P7, Canada

[b] Institut de Physique et Chimie des Matériaux de Strasbourg (IPCMS), UMR 7504, Université de Strasbourg, CNRS, 67034 Strasbourg, France

E-mail. kenneth.beyerlein@inrs.ca

**ABSTRACT**

Rapid and controllable reduction of graphene oxide (GO) remains a critical challenge for realizing its full technological potential. Here, we report the reduction of GO by a synergistic electron-beam–assisted single-pulse near-infrared (NIR) laser process. We find that the efficiency of 1064-nm laser photoreduction can be dramatically enhanced by prior exposure of the sample to a high-energy electron beam. Time-resolved electron energy-loss spectroscopy measured with a dynamic transmission electron microscope (DTEM) is used to quantitatively track the oxygen concentration evolution after subsequent NIR laser pulse irradiation. This finds an oxygen diffusivity for the process of $1.4 \pm 0.3 \times 10^{-9}$ $m^2/s$, which suggests oxygen is released by diffusion through defects and imperfections in the film. Electron beam irradiation is found to change the optical absorptivity of GO in the NIR region, and COMSOL simulations are carried out to estimate the thermal heating cycle resulting from the laser pulse. Structural characterization via selected-area electron diffraction (SAED) and high-resolution transmission electron microscopy (HRTEM) find localized restoration of $sp^2$ bonding accompanied by turbostratic disorder in the reduced GO. Together, these results point to a mechanism involving the creation of defects and vacancies produced by electron beam irradiation, which increases the efficiency of NIR light absorption and oxygen diffusion normal to the layers. This study demonstrates the important role of such defects in controlling the photochemistry of GO and its photoreduction kinetics.

**Keywords**: graphene oxide, near-infrared photoreduction, electron beam interaction, dynamic transmission electron microscope, electron energy loss spectroscopy

## 1. Introduction

Graphene oxide (GO), with its rich chemical landscape of oxygen-containing functional groups, has emerged as a uniquely versatile material for a wide range of applications[1-3]. Nonetheless, optimization of its transformation into reduced graphene oxide (rGO) is widely sought as a scalable means to produce materials with graphene-like properties, such as high conductivity and mechanical stability. Among various strategies, light and electron beam processing have gained attention as a controllable approach for probing and engineering GO at the nanoscale[4-6]. Localized irradiation of GO with light and electron beams enables its selective reduction, allowing direct patterning and writing of circuits and electronic devices within GO thin films[7-10]. While the diversity of oxygen groups in GO grants its tunable chemistry, it also complicates the reduction process[11]. The final properties of the material are sensitive to aspects such as the amount of remaining oxygen and the nature of defects in rGO[12]. A deeper understanding of the mechanisms involved in these reduction methods is necessary to improve manufacturing efficiency and quality of rGO devices.

The reduction of GO by light proceeds through a combination of photothermal and photochemical mechanisms. In photothermal reduction, light absorption leads to rapid local heating, which drives the sequential removal of oxygen functional groups across a broad temperature range (100–1200 °C)[13, 14]. In contrast, the photochemical pathway involves absorption of photons with energies in the range of ~2.4 to 4 eV, promoting electronic transitions between the π and π* states. These photogenerated electron-hole pairs participate in redox reactions, facilitating the removal of oxygen functionalities in GO[15-19]. For lower photon energy, photochemical and photothermal pathways operate simultaneously, creating a subtle interplay between light-induced bond activation and heat-mediated structural transformation. The balance between these mechanisms, strongly influenced by absorptivity, fluence, and pulse duration, governs the complexity of GO’s light-driven evolution and enables opportunities to tailor the chemical nature and properties of the resulting material[20]. Ultrafast laser studies can offer a way to disentangle the roles of photochemical and photothermal mechanisms by measuring the timescales on which photoinduced transformations occur[18]. In a recent study, Gengler et al.[21] employed femtosecond transient absorption spectroscopy to probe GO photoreduction in solution and proposed an ultrafast photoinduced chain reaction. Initiated by UV-induced photoionization of the solvent, the process released solvated electrons that rapidly drove reduction, with reaction dynamics unfolding in the picosecond regime. Further, Fatkulin et al.[22] revealed the dominant role of photochemical reactions in the laser-induced reduction of GO under visible light, challenging the widely held view that photothermal effects prevail. However, their study was confined to the spectral range of 405-633 nm, and the effects of longer wavelengths, such as 1064 nm, remain unexplored. Irradiation of GO with near-infrared (NIR) light has been touted as an efficient means for photothermal reduction[23, 24]. However, questions remain, as GO has poor absorptivity in the NIR, and significant reduction can require long exposure times of continuous and high-intensity pulsed lasers [24-27].

The modification of GO by electron beam irradiation is equally complex, as inelastic scattering events lead to atom displacements, electronic excitation, and ionization that trigger a cascade of electrochemical reactions in the material. The effects of electron kinetic energy and dose have been reported, revealing that sufficient electron exposure can partially reduce and alter its structural order. Low- and medium-energy irradiation (5–20 keV) doesn't lead to atom displacements but has been shown to modify the chemical composition of GO at doses approaching ~$10^{13}$ $e^{-}/nm^{2}$, leading to measurable changes in oxygen content and bonding configuration[28]. At significantly higher energies, Chen et al. demonstrated reduction of GO using a 5 MeV industrial accelerator at an absorbed dose of 500 kGy, where atom displacement damage predominates. XRD and XPS confirmed decreased interlayer spacing and an increase in C–C bonding, accompanied by a slight rise in structural disorder[29]. Similar trends were observed under 25 keV SEM irradiation, where fluences up to $10^{13}$ $e^{-}/cm^{2}$ modified GO over sub-centimeter areas and changed its bandgap[30]. High-energy transmission electron microscopy studies (300 keV) further showed that focused beams with current densities from ~0.5 to ~3.3 $mA/cm^{2}$ drive substantial oxygen loss and lattice restructuring under continuous or intermittent irradiation[5]. At the typical electron energies in transmission electron microscopy (100 – 300 keV), displacements of hydrogen, carbon, and oxygen atoms occur, in addition to electronic excitations and bond breaking[31]. However, the formation of rGO with a low oxygen content and a high degree of order has not been reported. Furthermore, little information exists on the local chemical changes happening through the irradiation process.

Driven by these outstanding questions concerning the effect of the electron beam and photoreduction processes, we carried out experiments on GO films using a dynamic transmission electron microscope (DTEM), combining high spatial and high temporal resolution. A DTEM enables in situ observation of rapid and irreversible structural, chemical, and electronic transformations in nanomaterials by single-shot pump-probe experiments[32-35]. Sinha et al. demonstrated the capability of time-resolved electron energy-loss (EELS) spectroscopy using a DTEM by capturing the fast reduction of nanocrystalline NiO following infrared laser pulses[36]. Using nanosecond electron pulses, they resolved the transformation of NiO into liquid nickel within microseconds, followed by recrystallization into the solid phase. Their results revealed a high-temperature, first-order reaction pathway distinct from that of bulk materials, highlighting how nanoscale kinetics and transient phases govern laser-induced chemical reactions.

In this study, we will use this unique instrument to study the complementary nature of electron and laser irradiation in the reduction of GO. Through in situ electron energy loss spectroscopy (EELS) measurements, we will show that separately electron beam and NIR light irradiation are not effective in reducing GO. However, when combined, the photoreduction efficiency is dramatically enhanced, allowing for localized reduction in a single nanosecond laser pulse. Furthermore, we will use the time-resolved ability of the DTEM to capture the progression of GO reduction in real time and quantify the kinetics of this transformation as it unfolds on nanosecond timescales. This combined with high-resolution characterization, thermodynamic simulations, and selected area electron diffraction (SAED), allows us to determine the mechanism for electron beam-assisted photoreduction of GO.

## 2. Experimental section

### 2.1. Synthesis of Graphene oxide (GO)

GO was synthesized via a modified Hummers' method[37, 38]. In a typical procedure, natural graphite flakes (1 g, 325 mesh, 99.8%, Alfa Aesar) were dispersed in a mixed acid solution of concentrated sulfuric acid ($H_2SO_4$, 23 mL, 95%, Sigma-Aldrich) and nitric acid ($HNO_3$, 6 mL, 70%, Sigma-Aldrich) and stirred for 1 h at 0 °C. Potassium permanganate ($KMnO_4$, 6 g, ACS grade, American Chemicals Ltd.) was then slowly added under constant stirring at 0 °C. The reaction mixture gradually transformed into a green slurry at approximately 35 °C and was subsequently stirred for an additional 3 h at 80 °C. Deionized water (60 mL, 18.2 MΩ cm, Millipore Milli-Q) was then added slowly to the suspension while maintaining the temperature at 80 °C, followed by further dilution with deionized water (140 mL) and hydrogen peroxide ($H_2O_2$, 20 mL, 30%, Sigma-Aldrich). The resulting suspension was centrifuged at 7000 rpm and washed repeatedly with deionized water until a neutral pH was achieved. The product was then freeze-dried for 48 h to obtain purified GO.

For transmission electron microscopy, a diluted GO suspension was drop-cast onto bare 2000 mesh Cu TEM grids (SPI Supplies, pitch: 13 $\mu$m, hole: 8 $\mu$m, bar: 5 $\mu$m). This grid was used to provide good mechanical support to the GO films, maximize heat conduction and avoid the detection of extraneous carbonaceous species during quantification. The samples were dried overnight in a sealed, controlled environment to prevent adsorption of moisture or other airborne contaminants.

### 2.2. Optical characterization of GO

To assess the optical response of GO under near-infrared excitation, we performed multiple analysis to determine its absorption coefficient. First UV–visible spectroscopy was performed on drop-cast GO films of 1 um thickness revealing a weak absorbance at 1064 nm of 0.1 (Figure SI3), which was later used to determine the linear absorption coefficient of the pristine sample before irradiation.

Kramers-Kronig analysis (KKA) of low-loss EELS spectra was also carried out before and after intense electron beam irradiation (Figures SI4 and SI5). In doing so, static EELS spectra were acquired with an energy resolution of 1–2 eV, as determined from the FWHM of the zero-loss peak (ZLP). Plural scattering effects were removed using the Fourier-log deconvolution method in Digital Micrograph, and the deconvolved spectra were smoothed using a low-pass numerical filter prior to KKA to extract the real and imaginary parts of the dielectric function ($\varepsilon_1$, $\varepsilon_2$), from which the absorption coefficient was derived.

### 2.3. In situ laser irradiation and DTEM measurements

Time-resolved experiments were carried out using a 200 keV DTEM[36, 39] (modified JEOL 2100) equipped with an electron energy-loss spectrometer (Gatan Enfinium) at IPCMS. This instrument is modified from a conventional TEM to be coupled with two pulsed lasers: one laser directly excites the specimen, while the second laser illuminates the photocathode to generate short electron pulses. In our study, the GO samples were irradiated with a single pulse from the fundamental output of a neodymium: yttrium aluminum garnate (Nd:YAG) laser with a duration of 7 ns, wavelength of 1064 nm, and a gaussian spot size of ~150 $\mu$m. Electron pulses were generated from the photocathode using the fifth harmonic (213 nm) of the fundamental emission. The timing of the electron pulse was electronically controlled to record single-shot EELS measurements after the laser excitation with nanosecond temporal accuracy. Spectra were acquired using an electron beam with a 100 nm probe diameter and an energy resolution of approximately 40 eV, which was sufficient for reliable quantitative analysis. Measurements at each time delay were repeated five times on different areas of the sample, and the resulting spectra were summed to enhance the signal-to-noise ratio. Each measured area was spaced sufficiently far apart to avoid being pre-irradiation by the laser and electron beams.

The post-irradiation EELS, imaging, and diffraction measurements presented later in this study were collected using the dynamic transmission electron microscope at INRS-EMT (modified JEOL JEM-2100 Plus). In this case, the microscope was operated in continuous thermionic emission mode at 200 kV and the sample was irradiated with an 11-ns, 1064-nm Nd:YAG laser pulses (Northrop Grumman) with the same fluence but a larger spot size of 400 µm. Electron energy-loss spectra were collected using a Gatan Quantum GIF system, with an objective aperture corresponding to a semi-collection angle of 12.7 mrad. Following in-situ laser reduction of GO, high-resolution TEM imaging was carried out at the McGill Facility for Electron Microscopy Research using a Thermo Scientific Talos F200X G2 (S)TEM.

**3. Results and Discussion**

We begin by investigating the extent to which the sample composition changes due to intense electron beam irradiation. Core-loss spectra were collected on a fresh area of the sample using a spread beam before and after irradiating it with a tightly focused beam at 200 keV electron energy for 30 seconds. This electron beam exposure corresponds to an electron dose of approximately $6 \times 10^7$ $e^-/nm^2$. As seen in Fig. 1a, the core-loss EELS spectra of the C and O K-edges of GO exhibited distinct changes after this level of irradiation. A C/O ratio of 2.1 (68 ± 3 at. % C, 31 ± 2 at. % O) for the pristine GO sample was found by EELS elemental quantification performed using the Gatan Digital Micrograph software. After electron beam irradiation, a measurable decrease in the O K-edge intensity was evident, and subsequent quantification found this electron dose to result in the removal of approximately 11 ± 1 at. % of oxygen. Concurrently, a broad pre-edge feature emerged at the O K-edge (Figure SI1), which is characteristic of oxygen in vacancies, epoxide conversion to –OH, or carbonyl species[40, 41]. This spectral evolution provides direct evidence of electron beam-induced modification of the C–O chemistry in GO, suggesting that the electron beam facilitates localized partial reduction and structural defects such as vacancies within the oxygenated carbon network. Previous studies have shown that high-energy electrons can readily break C–O bonds and promote the desorption of oxygen-containing functional groups, even under high-vacuum conditions, due to a combination of knock-on displacements of atoms and radiolysis[5, 29]. EELS spectra were then measured on a fresh area of the sample before and after irradiation with a 7-nanosecond laser pulse (1064 nm) and fluence of 100 mJ/cm$^2$, showing no change (Fig. 1b). Even upon increasing the number of pulses, the reduction with remained negligible, indicating that the laser energy was insufficient to overcome the bond dissociation threshold of oxygen-containing groups in GO.

However, localized removal of oxygen from GO was reproducibly found after exposing an area of the sample previously irradiated by the intense electron beam to a laser pulse. As shown in Fig.1c, EELS spectra acquired before and after irradiating the sample with an electron dose of 6x10$^7$ e$^-$/nm$^2$ followed by illumination with a NIR pulse reveal a pronounced change in the elemental composition of the material. Notably, the combined action of the electron beam and laser pulse (100 mJ/cm$^2$) reduced the sample to the point that the O K-edge was no longer detectable over an area of approximately 2 µm (Fig. 1c). Furthermore, the C K-edge spectra taken in the same region and shown in Fig. 1d exhibits a distinct emergence of the π* resonance alongside the σ* edge, indicative of a localized graphitic lattice with a coherent π electron system. The disappearance of the O K-edge confirms that the synergistic action of electron–photon excitation at this dose and fluence threshold enables localized and effective reduction of GO within a single pulse event (Fig. 1e). An electron dose in this range is crucial, as no distinguishable change in the sample composition was found when pre-irradiating with a lower dose of 9.35 x 10$^4$ e$^-$/nm$^2$ (Figure SI2).

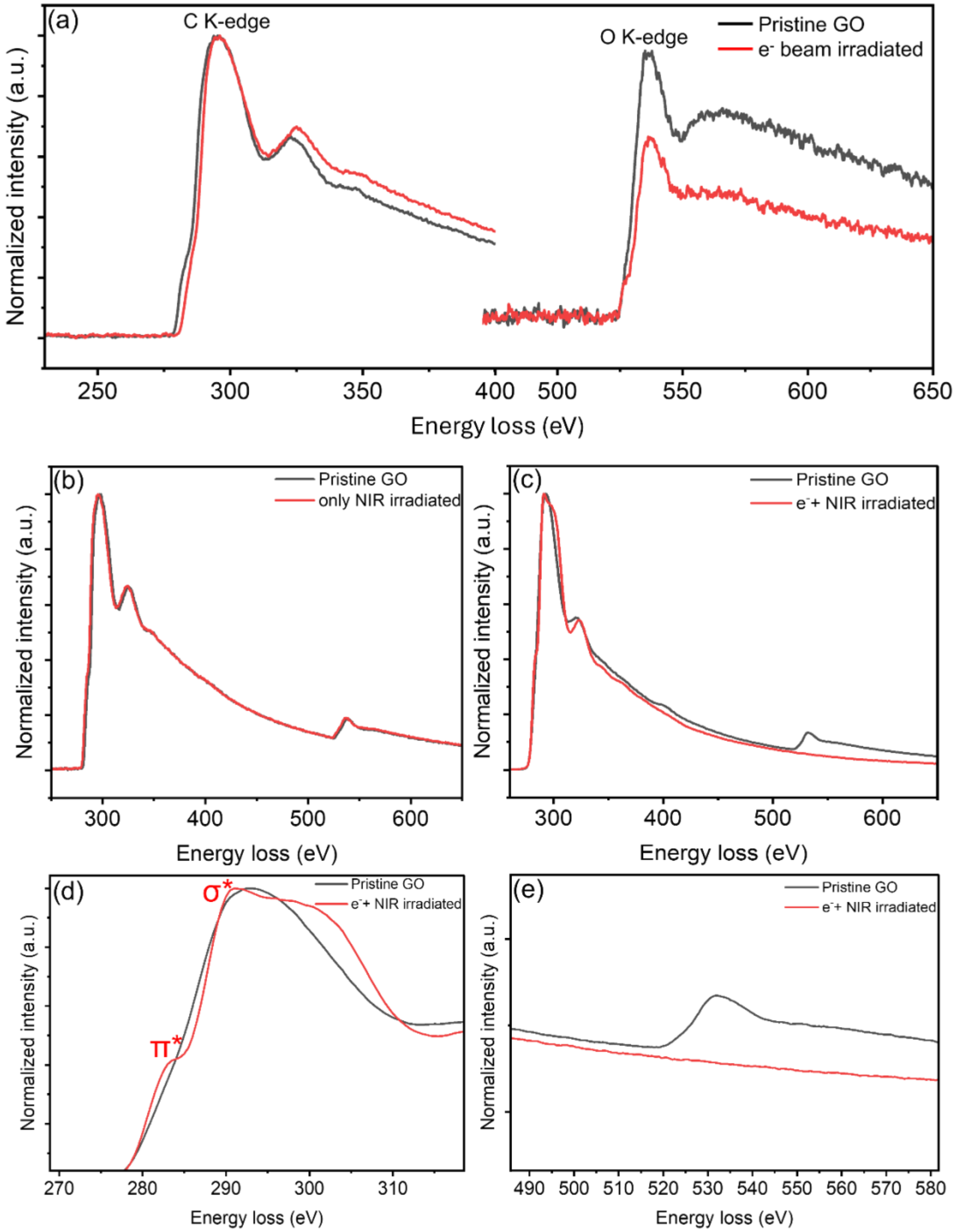

**Fig. 1.** EELS spectra of GO showing the C K-edge and O K-edge features under different irradiation conditions. (a) Spectra acquired before and after exposure to a 200 keV electron beam with a dose of $6x10^7$ $e^{-}/nm^2$. (b) Spectra collected following illumination with a single NIR laser

pulse. (c) Spectra obtained before and after irradiation of the electron beam followed by a single NIR laser pulse (100 mJ/cm$^2$). (d) A zoomed in view of the C K-edge from (c) shows the emergence of the π* resonance alongside the σ* edge, (e) while any measurable sign of the O K-edge is completely removed, evidencing localized reduction of GO.

To capture the temporal evolution of this enhanced NIR GO reduction, we performed a series of single-shot time-resolved EELS measurements. Details of the procedure are described in the methods section. The degree of reduction was determined by fitting the O K-edge intensity as a function of pump-probe delay. Fig. 2a presents EELS spectra acquired for delays ranging from 10 ns to 2 μs, alongside the reference spectrum representing pristine GO taken before laser irradiation. Given the inherently low signal-to-noise ratio in single-shot EELS, each spectrum represents an average over multiple measurements, each taken in single-shot mode in different regions of the sample to ensure reliability. The data are overlaid with a fit of the C K-edge modelled as a smooth decaying edge to highlight the oxygen edge signal at ~532 eV. The O K-edges after normalization to the C edge and background subtraction are shown in Fig. 2b. A pronounced decrease in the O K-edge intensity is evident as early as 10 ns after the NIR pulse, indicating the rapid onset of deoxygenation.

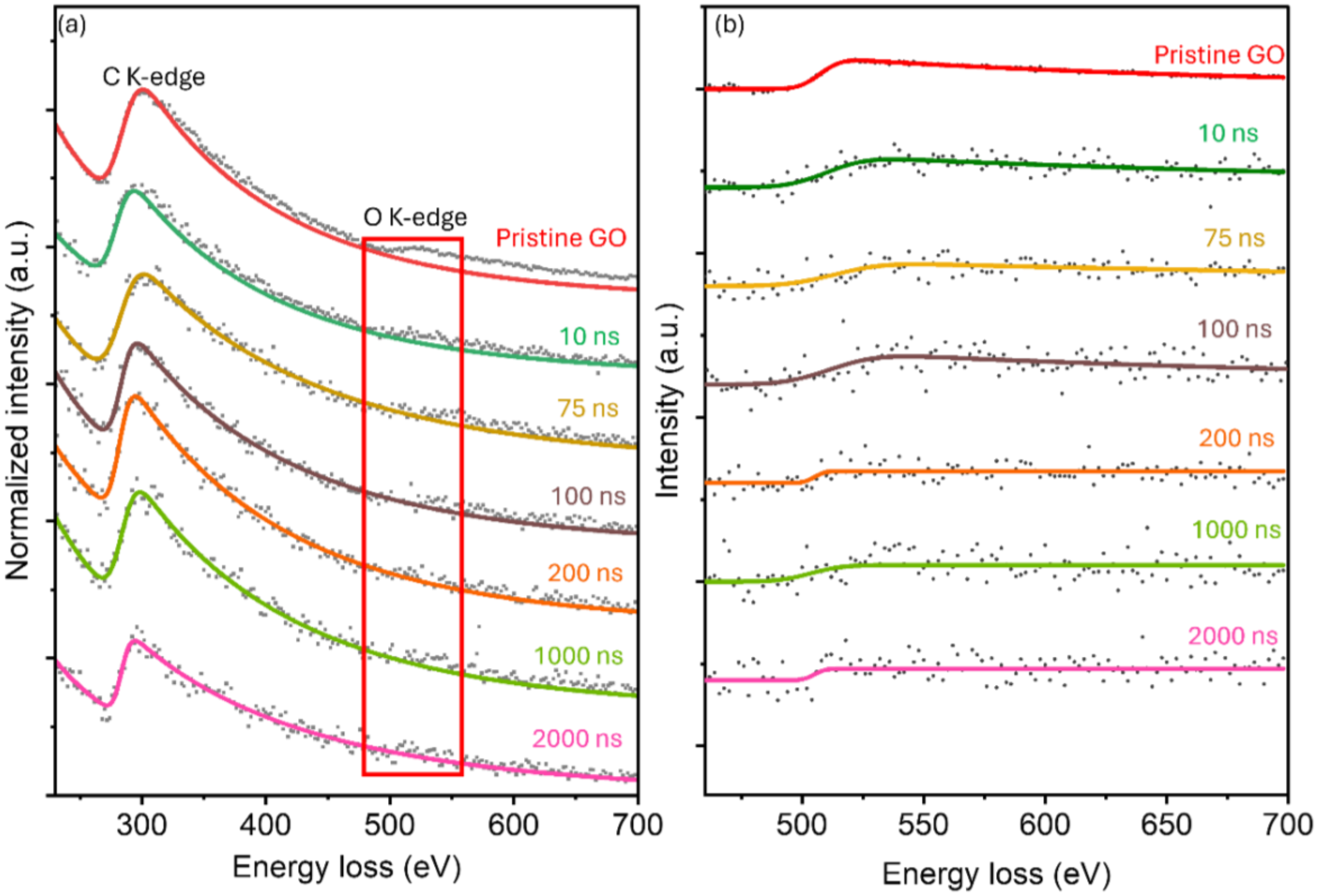

**Fig. 2.** Time-resolved EELS of GO photoreduction. (a) Core-loss spectra in the range of the C K-edge and O K-edge acquired using single 7-ns electron probe pulses at varying delay times following excitation by an NIR laser pulse. (b) The O K-edge data after background subtraction were fit by an error function with a decreasing amplitude with time delay, indicating the evolution of oxygen removal from the film.

The acquired EELS spectra were analyzed to determine characteristic parameters of the O K-edge. First, the background coming from the tail of the low-loss plasmon peak in the range of 225 eV to 260 eV was modeled as a power-law function and subtracted from the spectrum. Then the spectrum was normalized to the maximum value around 300 eV corresponding to the height of the carbon edge, which is shifted from the normal value of 285 eV due to broadening caused by the 40-eV energy resolution of the microscope in photoemission mode. The background of the O K-edge after this normalization was also modelled by a power law in the range of 450 eV to 500 eV and subtracted to obtain spectra like those shown in Fig. 2b.  The resulting O K-edge was fit as an exponentially decaying error function following

$$I(E) = O_x \exp\left(\frac{-(E-\mu)}{\tau_E}\right)\left[1 + \text{erf}\left(\frac{(E-\mu)}{\sigma}\right)\right]/2, \tag{1}$$

where $\mu$ and $\sigma$ denote the position and broadening of the edge, while $O_x$ and $\tau_E$ denote its amplitude and energy loss decay constant. Nonlinear least squares fitting was performed using the Python package LMFIT, and uncertainties in the best-fit values were obtained from standard errors of the fit parameters. The fits shown in Fig. 2b highlight the decreasing oxygen edge amplitude with time and an apparent slight shift in the oxygen edge position for times after 100 ns. However, further investigation found that this shift is largely due to uncertainty in the fitting caused by a decreased signal level and inherent noise in the spectrum. Therefore, the investigation focused on the trend of the oxygen edge amplitude from single-shot spectra recorded as a function of time delay, denoted as $O_{ss}(t)$. Spectra collected before and after laser irradiation on each spot were also averaged, normalized, and fit to determine the oxygen edge amplitudes, $O_b$ and $O_a$ respectively. The relative change in the oxygen edge amplitude given by $(O_{ss}(t) - O_a)/(O_b - O_a)$ was then calculated to track the progress of oxygen removal over time and is plotted in Fig. 3. This quantity was used to account for variations in local oxygen concentration as well as slight changes in photoreduction efficacy caused by differences in thickness and light absorption in different areas.

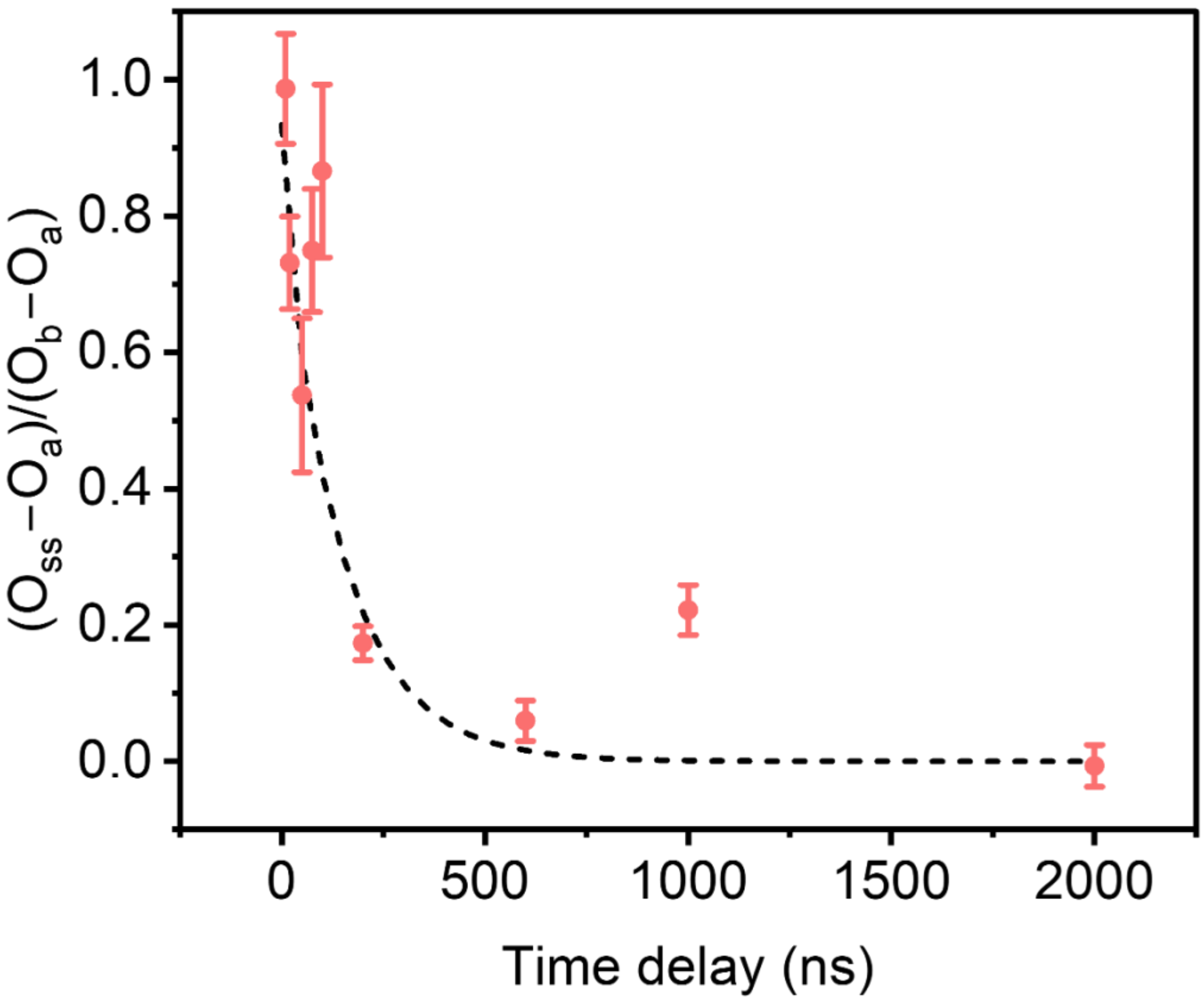

**Fig. 3.** The relative change in the oxygen edge amplitude is defined as $O_{ss}(t) - O_a/O_b - O_a$ is plotted as a function of time delay. The error bars on the data were determined from uncertainty in fits to the O K-edge modelled as an error function. The best-fit model following Eqn. 3 is also shown as a dashed line.

A simplistic model for the evolution of oxygen in the laser-irradiated GO thin film was derived based on the theory of diffusion in solids. The laser-induced removal of oxygen from the thin GO membrane in our experiment can be thought of as an extreme case of a thin film outgassing in vacuum. An analytical solution describing the spatio-temporal evolution of the gaseous species concentration in the film can be found from Fick's second law, assuming that the concentration outside of the film is zero. This boundary condition results in a function of the form[42].

$$c(x,t) = \frac{4c_0}{\pi}\sum_{j=0}^{\infty}\frac{1}{2j+1}\sin\left[(2j+1)\pi\frac{x}{w}\right]\exp\left[-tD\frac{(2j+1)^2\pi^2}{w^2}\right]. \quad (2)$$

The formulation of the concentration distribution as an infinite Fourier series comes from the necessity to describe the initial rectangular spatial distribution of the species concentration in the film, having a sharp edge on the top and bottom surfaces and a near-constant value ($c_0$) through the thickness ($w$). The gradual decrease of the species concentration with time is captured by the decaying exponential functions, which is dependent on the diffusivity of the species in the film ($D$). To compare with EELS measurements, which measure the average concentration through the film, this equation was integrated between $x = 0$ and $x = w$, and normalized to find

$$\bar{c}(t) = \frac{1}{wc_0}\int_0^w c(x,t)\,\mathrm{d}x = \frac{8}{\pi^2}\sum_{j=0}^{\infty}\frac{1}{(2j+1)^2}\exp\left[-tD\,\frac{(2j+1)^2\pi^2}{w^2}\right]. \quad (3)$$

It is worth noting that the exponential damping function is dependent on the square of the order ($j$), which means that as time progresses, higher-order terms are damped faster and contribute considerably less to the concentration distribution. Plots of Eqn. 3 with increasing orders were used to determine that including terms up to the 5$^{th}$ order was sufficient to capture the time dependence of the trend over the measurement range.

The data for the relative change in the oxygen edge amplitude shown in Fig. 3 were then modeled using Eqn. 3, assuming a film thickness of 46 nm and $j_{max}$= 5. The best fit overlaid on the data corresponds to a diffusivity value, *D*, of 1.4 ± 0.3 x $10^{-9}$ $m^2/s$. It is important to note that this model only considers the diffusion of oxygen in the direction normal to the film surface, and not within the plane. As simulations will later show, the sample is rapidly heated by the laser, achieving a peak temperature well above room temperature within just 10 nanoseconds. This model then assumes that dissociation of the oxygen from the carbon network and formation of stable oxygen dimer molecules happens rapidly once this temperature is reached. The process of oxygen diffusion may involve further intermediate reactions transforming oxygen species as well as repeated adsorption and desorption during transport; however, the temporal resolution and data quality of the experiment do not allow for the discrimination of these transient states.

This diffusivity obtained in this experiment can be compared with values obtained through gas pressure differential measurements on GO films. The diffusivity of nitrogen gas through carbon-based films (thickness > 1 $\mu$m) such as graphite, rGO, and GO has been measured to be in the range of 3 x $10^{-8}$ $m^2/s$ at room temperature[43, 44]. This value is also consistent with intralayer diffusivity coefficients found from molecular dynamics simulations[45]. However, compact GO films with a thickness of 33 nm have been measured to exhibit nitrogen and oxygen diffusivity in the range of 3 x $10^{-15}$ $m^2/s$. This large discrepancy in the literature is explained by increased density and flatness of the GO flakes of the later measurements, preventing the diffusion of oxygen through cracks and gaps between flakes. The diffusivity of 1.4 ± 0.3 x $10^{-9}$ $m^2/s$ measured here by time-resolved EELS is therefore consistent with the diffusivities reported for less dense and thicker films[43, 44]. This suggests that oxygen is primarily diffusing through gaps and defects in the film during the laser-induced photoreduction process.

It should be noted that the temperature of the thin film will be significantly higher than room temperature as the oxygen is diffusing out of the film, which can also affect the oxygen diffusivity. We calculated the temperature evolution in the film as a function of time using the heat transfer module of COMSOL Multiphysics (Fig. 4). The dimensions of the film in the simulation are shown in the figure and chosen to match the experimental configuration. The effect of laser heating was simulated as 7-ns temporal heat pulse with a two-dimensional gaussian spatial distribution following the profile of the laser spot at IPCMS. The boundary of the film was held constant at room temperature to simulate the rapid heat conduction into the surrounding copper grid. An accurate estimation of the temperature requires precise knowledge of the material's physical properties, which can strongly depend on its preparation method. As discussed in the experimental section, we measured the linear absorption coefficient of our GO samples using two complementary approaches: direct UV–visible spectroscopy and KKA of low-loss EELS spectra. From the measured absorbance of pristine GO, using the film thickness determined by SEM of its cross-section (1 μm), we obtained an absorption coefficient of approximately $2\times10^5$ $m^{-1}$ (Figure SI3). To ensure a consistent comparison, the absorption coefficient of pristine GO was also determined via EELS/KKA, yielding a value of ~$1.8 \times 10^5$ $m^{-1}$ (Figure SI4), in good agreement with the UV–Vis-derived value (~$2 \times 10^5$ $m^{-1}$). However, EELS/KKA analysis of different regions of the sample after e-beam irradiation found an increased absorption coefficient of ~$1 \times 10^6$ $m^{-1}$ (Figure SI5). We note that at the electron energy loss probed (~1.16 eV) is at the boundary of the energy resolution limit of our static EELS measurements (1–2 eV, determined from the ZLP FWHM), and residual ZLP tail intensity may introduce additional uncertainty into the absolute values reported in this region. Nonetheless, our measurements support an increase in the absorption coefficient upon irradiation as these data were measured and evaluated consistently across the same spectral range for pristine and irradiated samples. Namely, our analysis finds that the absorption at 1064nm increases by 5.5x after electron-beam irradiation, which partially explains the enhanced photoreduction efficiency. These values were then incorporated into the COMSOL Multiphysics simulations to estimate the laser-induced temperature rise under a single NIR pulse. Assuming the parameters of pristine GO, a modest peak temperature of 68 °C was found, whereas the parameters of e-beam–modified GO resulted in a peak temperature of 206 °C (Fig. 4). A parametric simulation study was conducted and found that this temperature is predominantly determined by the absorption coefficient and largely insensitive to the thickness and heat conductivity of the material (Figure SI6). It is also worth noting that the primary channel for cooling is heat conduction in the plane of the film, as the sample is suspended in a vacuum. Estimates of the cooling rate from the slope of the curve suggest a cooling time of nearly 2 milliseconds for the sample to return to room temperature. Therefore, the sample can be assumed to be at a constant temperature of roughly 206 °C for the duration of the reduction process. As previously mentioned, temperatures near 1200 °C have been reported to be necessary for thermal reduction of GO[46]. The lower temperatures obtained from our simulations suggest that electron beam irradiation is not only increasing the absorptivity but also causing chemical and structural changes to the material to decrease this threshold.

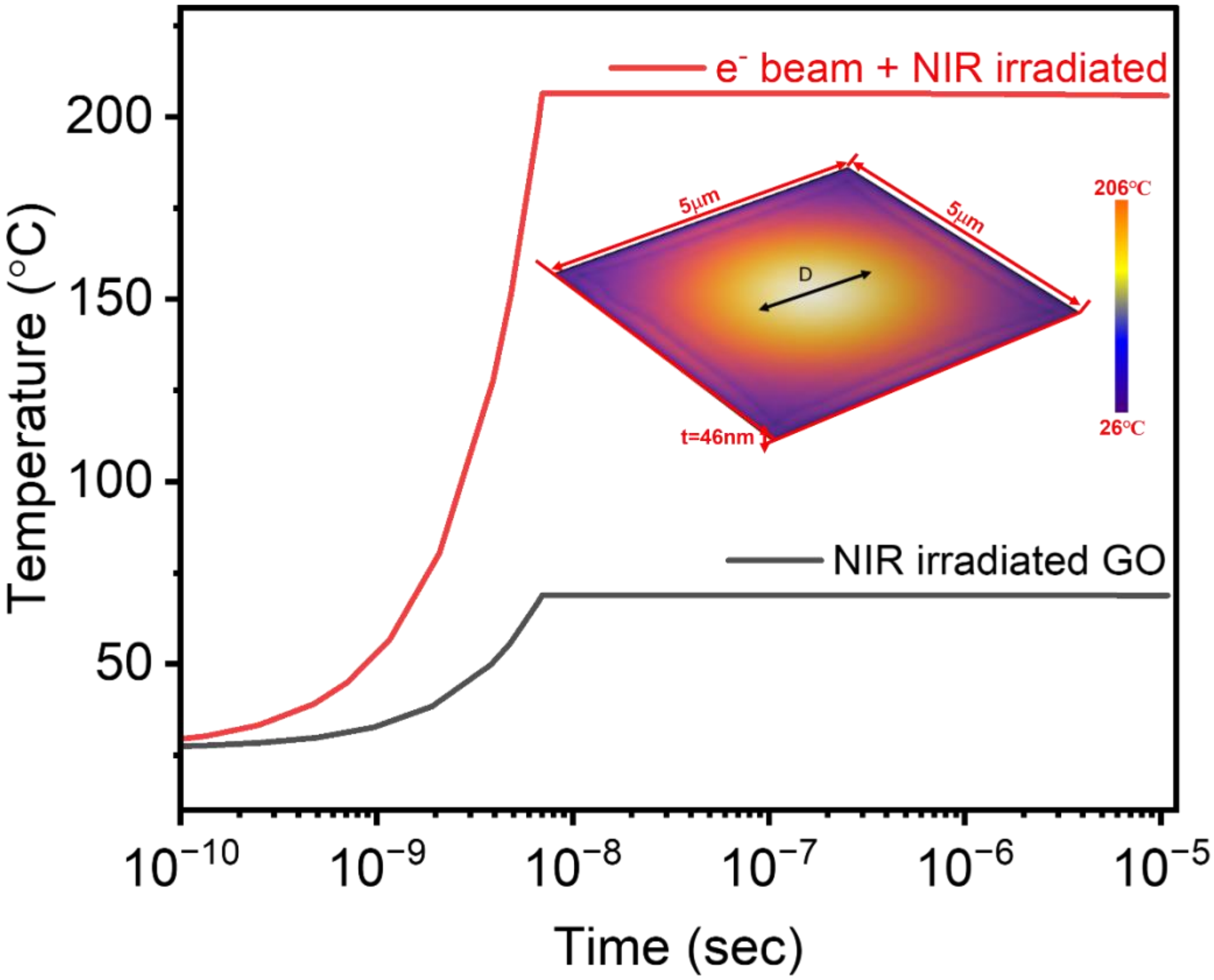

**Fig. 4.** Simulated temporal temperature profiles of pristine GO and electron-beam-irradiated GO following a single 1064-nm, NIR pulse obtained from COMSOL Multiphysics simulations. The irradiated material exhibits an enhanced temperature rise due to the increased NIR absorption coefficient that was taken from Kramers-Kronig analysis of low-loss EELS measurements of the sample after electron beam irradiation.

We also measured SAED images before and after electron beam and single-pulse NIR laser irradiation to study the ordering of the carbon bonding network happening in the process (Fig. 5(a-c)) and extracted azimuthally averaged radial intensity line profiles from each pattern for quantitative comparison (Fig. 5d). Prior to irradiation, the diffraction pattern exhibited diffuse rings, characteristic of a polycrystalline material lacking long-range order or preferred stacking between GO layers (Fig. 5a). Exposure of GO to the electron beam produced slight modifications in its diffraction signature, reflecting the well-established beam-induced deoxygenation and structural rearrangement of GO observed in prior studies[47]. Following electron beam irradiation, the SAED pattern shows a modest suppression of the (101) diffraction ring accompanied by a relative enhancement of the (112) reflection (Fig. 5b). Notably, the positions of the (101) and (112) reflections remain largely unchanged between pristine GO and electron-beam–irradiated GO, indicating preservation of the underlying in-plane lattice spacing. This evolution indicates that oxygen removal occurs in a spatially inhomogeneous manner, leading to disruption of the local carbon framework rather than the development of coherent graphitic ordering. Rather than driving the GO film toward hexagonal stacking, electron-beam exposure predominantly induces vacancy-type defects, bond scission, and nanoscale lattice distortions, which collectively weaken the periodicity associated with the (101) spacing[48]. Consistent with this interpretation, high-resolution TEM images of the pristine GO and electron-irradiated region do not reveal the emergence of crystalline domains, but instead show distorted, amorphous sheets (Figure SI7 and SI8, respectively).

The diffraction pattern changed significantly after high intensity electron beam irradiation followed by irradiation by a nanosecond NIR pulse. The (101) ring near ~5 $nm^{-1}$ increased in intensity, accompanied by the emergence of a weak shoulder assignable to the (102) plane, which is attributed to short-range interlayer ordering induced by heterogeneous, rapid laser reduction, rather than long-range graphitization[49]. In contrast, the (112) ring near 8.5 $nm^{-1}$ decreased in intensity and became more diffuse (Fig. 5c), concurrent with the appearance of the (110) reflection. This opposite trend suggests that the NIR pulse produces short-range $sp^2$ domains, likely through thermally assisted carbon–carbon bond reorganization and the removal of residual oxygen functional groups, while not restoring long-range in-plane coherence[50], as independently supported by EELS measurements (Fig. 1d). Further, the (112) reflection broadens, and the (110) peak becomes discernible, accompanied by the appearance of a faint outer ring near ~9.7 $nm^{-1}$, indexed to the (114) plane (Fig. 5d). In contrast, the combined electron-beam and single-pulse NIR sequence enables localized $sp^2$ reconstruction while maintaining an overall turbostratic structure[51]. Further, SAED patterns extending to the (002) reflection region were obtained for pristine GO and $e^-$ beam+NIR-irradiated rGO (Figure SI10). The diffuse ring observed in pristine GO sharpens into a more defined, spotty ring after irradiation, with a measured d-spacing of ~0.353 nm, confirming the emergence of local graphitic (002) stacking order consistent with the HRTEM observations.

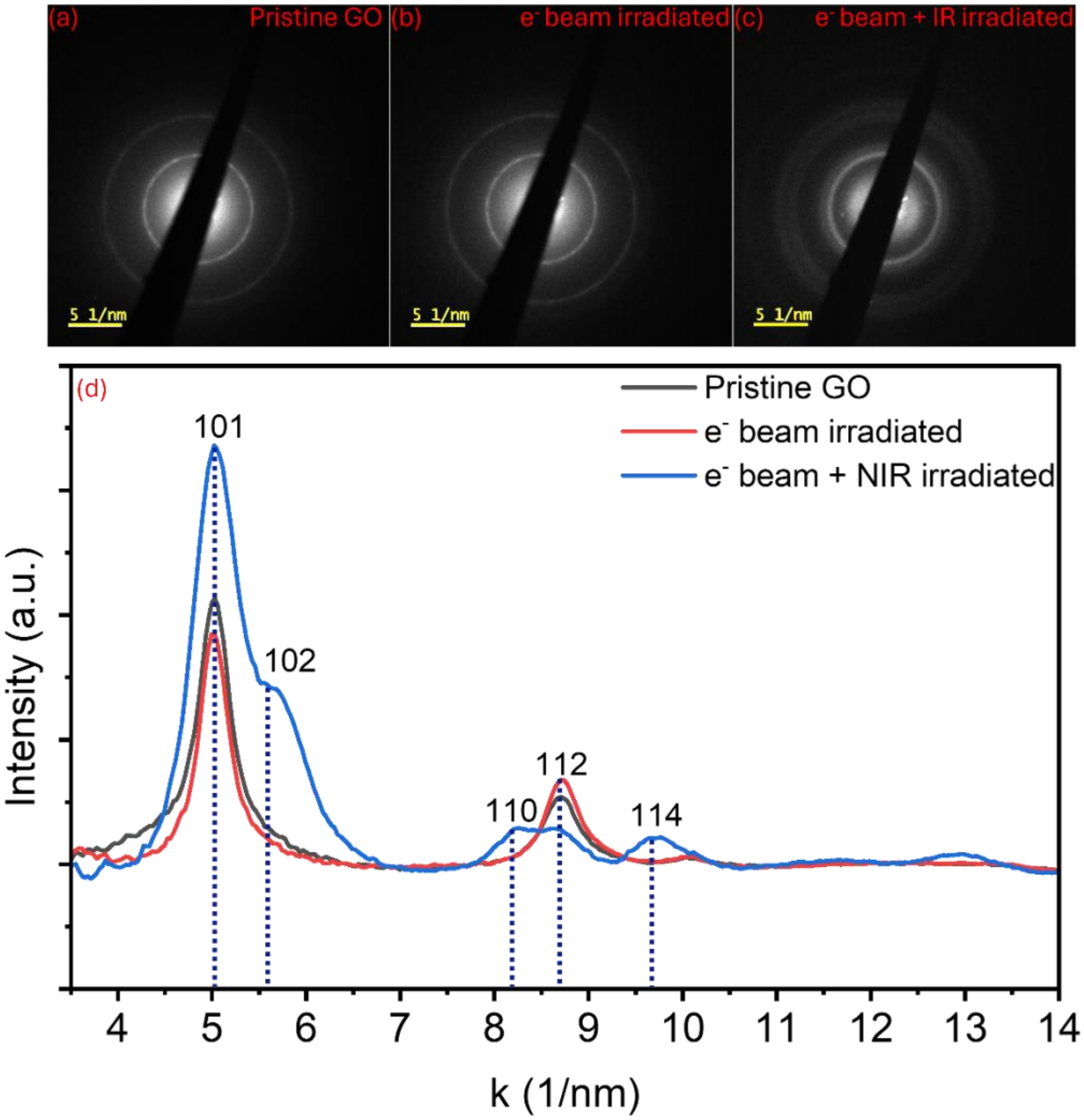

**Fig. 5.** Evolution of GO structure after electron-beam exposure and electron beam + single NIR pulse synergistic excitation. Selected-area electron diffraction patterns for (a) pristine GO, (b) GO after electron-beam irradiation, and (c) GO subjected to combined electron-beam and single 1064-nm NIR pulse excitation. (d) Azimuthally averaged radial intensity profiles highlight the distinct evolution of the (101), (102), (112), (110), (114), and higher-order features, revealing beam-induced disorder followed by e-beam and single NIR pulse-driven partial reconstruction of short-range $sp^2$ order.

Moreover, high-resolution TEM (HRTEM) imaging of pristine GO reveals a predominantly disordered structure, with no discernible lattice fringes (Figure SI7). While the region subjected to combined electron-beam and NIR-pulse irradiation reveals well-defined lattice fringes characteristic of locally ordered graphitic domains (Fig. 6a). Analysis of these fringes yields an interplanar spacing of approximately 0.34 nm, corresponding to the $d_{002}$ spacing of rGO[52] (Fig. 6b). However, it is also noteworthy that only a few small, ordered domains were found in the photoreduced area. This suggests that while irradiation with a single NIR pulse was able to remove the oxygen, it did not complete the graphitization process producing long range order in the material. To confirm the origin of the $d_{002}$ lattice spacing, additional HRTEM measurements were performed on sheet-like rGO regions (Figure SI11), yielding $d_{002}$ values of 0.368 nm and 0.38 nm, consistent with turbostratic graphitic stacking.

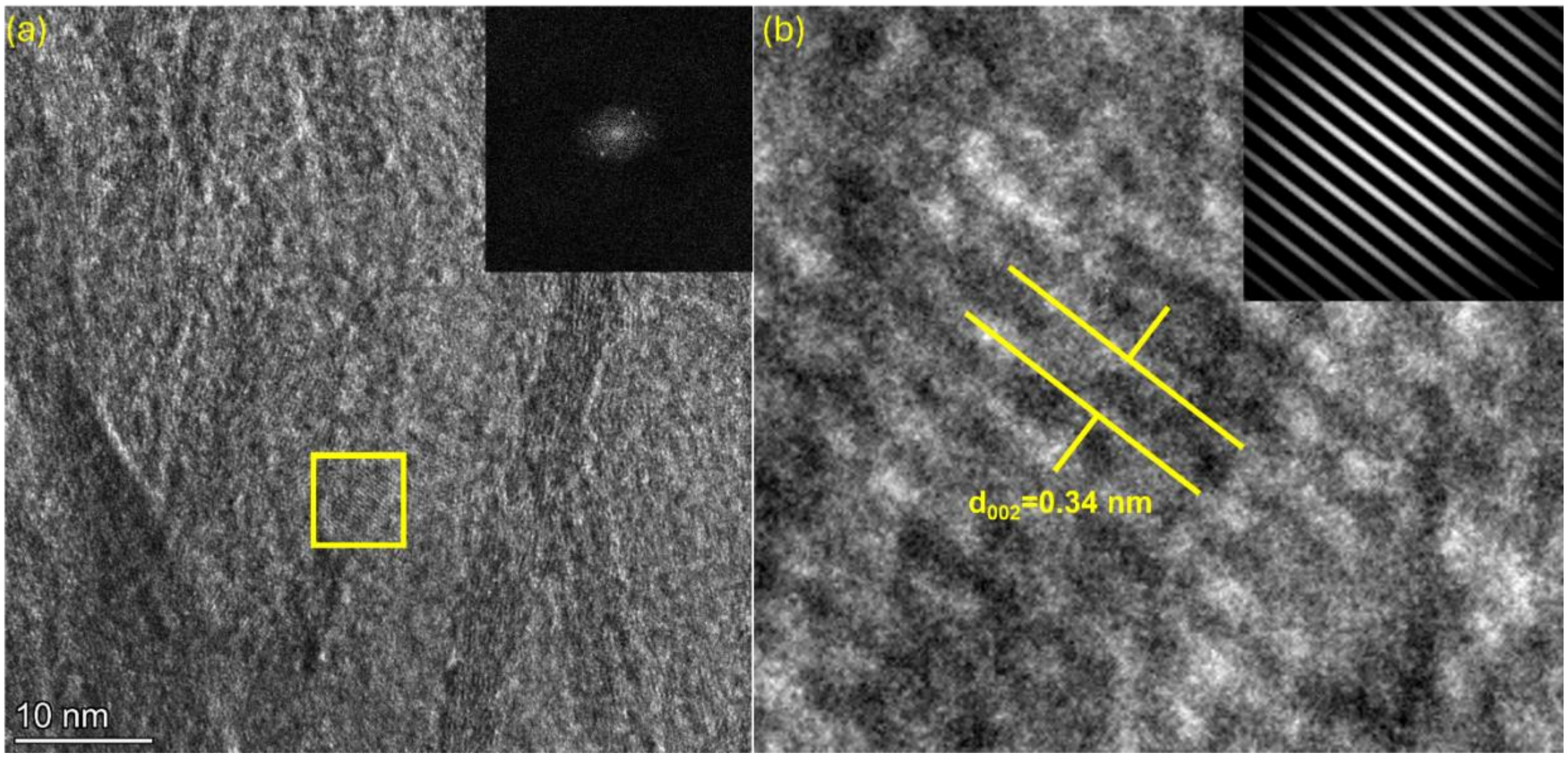

**Fig. 6.** HRTEM analysis of rGO after irradiation by an electron beam and a single NIR pulse. (a) High-resolution TEM image of the irradiated rGO region - the outlined rectangle indicates the area selected for detailed structural analysis, with the inset showing the fast Fourier transform (FFT). (b) Corresponding lattice-fringe image extracted from the cropped region, with the inset showing the inverse fast Fourier transform (IFFT) pattern generated from ImageJ analysis of panel (a), highlighting the local crystallographic ordering induced by the combined irradiation.

To accurately interpret the observed results, it is essential to consider the possible underlying mechanisms that could drive the reduction of GO. Given the intrinsically weak NIR absorption of GO ($2\times10^5$ m$^{-1}$) in comparison to pure graphene (~$10^7$ m$^{-1}$)[53, 54], no measurable reduction occurs under nanosecond NIR irradiation alone, making a purely photothermal pathway unlikely. COMSOL simulations incorporating the experimentally determined absorption coefficient, as well as the density, heat capacity, and thermal conductivity of GO, show that a single 7-ns NIR pulse (100 mJ/cm$^2$) increases the temperature only marginally (Fig. 4), far below the thresholds reported for thermally driven deoxygenation[55]. A second possible pathway is direct photochemical bond scission; however, the photon energy associated with NIR light used here (1.16 eV) lies well below the energy necessary to cleave epoxy groups (3.2eV) or weakly bound hydroxyl groups (1.5 eV), rendering photochemical removal of oxygen functionalities highly improbable. Taken together, these considerations indicate that neither thermal nor direct photochemical processes can account for the reduction observed in our experiments.

Instead, we propose that electron beam irradiation enables the photoreduction of GO with NIR light by creating defects and atomic vacancies in the GO sheets through a mechanism illustrated in Fig. 7. HRTEM experiments have shown that irradiation of GO by an intense 200 keV electron beam breaks carbon and oxygen bonds through radiolysis and knock-on mechanisms[31]. Even holes of a few nanometers in diameter have been produced in GO under somewhat higher electron doses[5]. Our observation of structural defects formed in the GO films by the electron beam was complicated by the thickness and amorphous nature of the initial material (Figure SI8). Nevertheless, an estimate of their concentration can be calculated assuming a similar picture to that of graphene. The number of vacancies created by electron irradiation in a graphitic lattice can be calculated from the number of displacements *(p)* per atom which is the product $p = \sigma j$ of the scattering cross-section *($\sigma$)* and the electron beam current density *(j)*. At an electron energy of 200 keV, assuming a displacement cross-section of $\sigma = 10^{-27}$ m$^2$ [56] and a beam dose of $6x10^7$ e$^-$ nm$^{-2}$, we obtain $p$ = 0.2 displacements per atom during an irradiation period of 30 s. However, it must be considered that not each atom displacement leads to a persistent defect, so the overall vacancy concentration in the irradiated region might be lower than 20%.

The creation of carbon vacancies has a twofold effect on improving the efficiency of GO photoreduction with NIR light. First, it increases the optical absorptivity of the material in the NIR regime. The high concentration of such vacancies present in laser-induced graphene has been shown to facilitate efficient absorption and annealing using femtosecond NIR laser pulses[57]. Simulations of pure graphene have shown that vacancy densities of 12% can change the absorbance by nearly an order of magnitude[58]. As our measurements and temperature simulations have shown a similar increase in absorbance, we assume a vacancy density in the same range.

Secondly, the creation of vacancies and defects improves the diffusion of oxygen through the stacked sheets of GO. Benzene rings which make up a perfect graphene lattice are in fact impermeable to any species at all realistic temperatures. Therefore, possible diffusion pathways for oxygen removal are (1) the space between the layers where lateral diffusion can occur, or (2) voids in the perfect lattice permitting diffusion in the normal direction through the layer. Lateral diffusion of oxygen is not expected to be a dominant pathway because of the strong interaction of the diffusing oxygen groups with the carbon layers[59-61]. Furthermore, the flake size of 2-5μm in the present samples would make an efficient escape of O atoms by this pathway unlikely. Diffusion normal to the planes (scenario 2) can occur by the penetration of O atoms through hole-like defects in the layers or by the temporary removal of a C atom in the layer by an O atom (exchange mechanism). However, the latter should not be related to the presence of radiation defects in the layer and can thus be excluded. The penetration of even heavy atoms through vacancies in spherical onion-like graphitic particles[62] or carbon nanotubes[63] has already been observed. Larger non-hexagonal rings such as heptagons, octagons, or nonagons are known from irradiation studies of graphene[31] and should facilitate the penetration of the layers. Furthermore, as previously discussed, our measured oxygen diffusivity of $1.4 \pm 0.3 \times 10^{-9}$ $m^2/s$ is consistent with values obtained from samples where diffusion between flakes and through cracks is dominant[43, 44].

We would like to add that TEM imaging of the affected area suggests that this rapid removal of oxygen from within the film can lead to the rupture of the surface and mass loss in the irradiated area (Figure SI9). It is important to note that the morphological rupturing observed in the irradiated region is a spatially confined phenomenon, limited to a nanoscale area, while a much larger micron sized area is reduced. As evident from Figure SI11, the dominant outcome of the combined electron-beam and NIR laser irradiation is the formation of structurally intact reduced graphene oxide sheets. HRTEM conducted on sheets located on the edge of the damaged region resolved a (002) lattice fringe spacing consistent with short-range graphenic interlayer stacking. Furthermore, the SAED measurements reflect a short-range interlayer ordering occurring across the bulk of the irradiated film. Therefore, in the final stage of the transformation, we find that the heat generated by the laser irradiation begins to heal the defect network and start the growth of local graphitic order in the material. This effect is marginal in the case of a single pulse, as we estimate the temperature of the system to be 206 °C upon laser irradiation and requires ~ 2 ms to cool back to room temperature (Fig. 4). However, it is enough to form small ordered graphitic domains in the affected area (Fig. 6). It also indicates that longer heating time at ideally higher temperatures is needed to complete the transformation and produce long-range order.

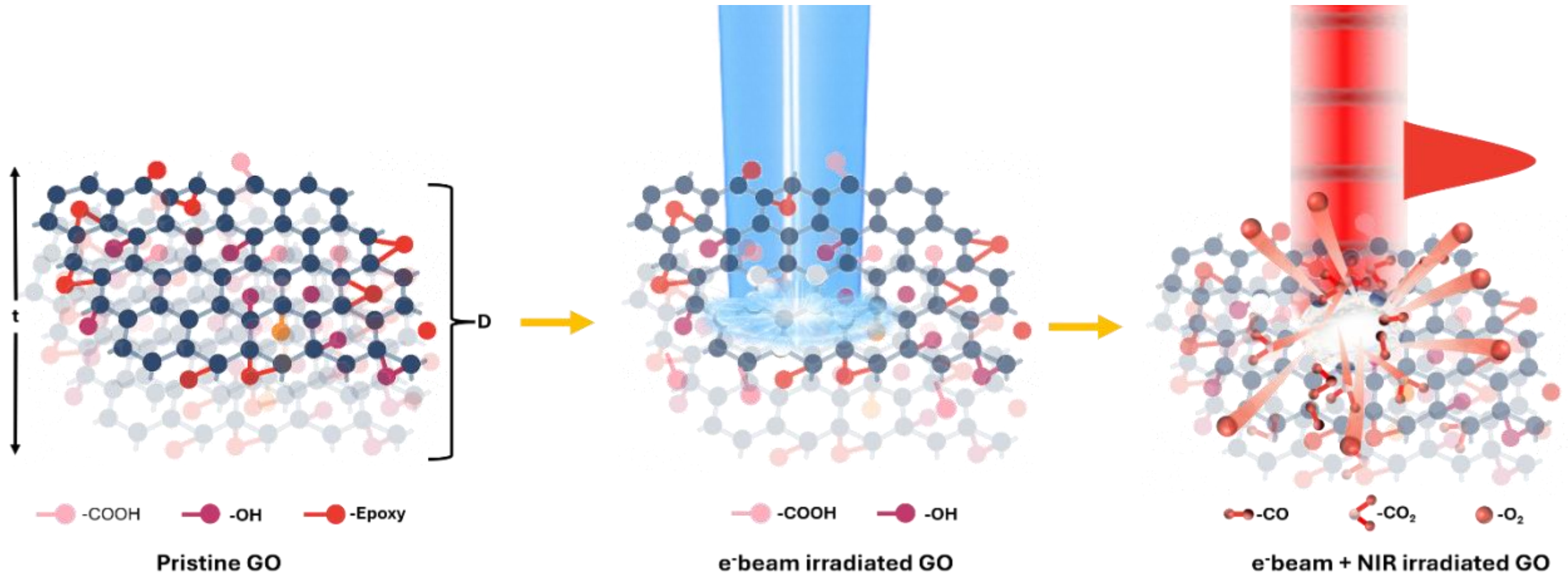

**Fig. 7.** Illustration of the proposed mechanism for electron-beam-assisted NIR laser photoreduction of GO.

## 4. Conclusions

Our work demonstrates the localized removal of oxygen from multilayer GO within hundreds of nanoseconds following a single NIR pulse can be achieved after pre-exposure of the sample to a high-energy electron beam. In the process, we uncover a synergistic photo-radiolytic reduction mechanism beyond conventional photothermal effects. We suggest that irradiation with an electron beam dose in the range of $6x10^7$ $e^-/nm^2$ produced a significant fraction of vacancies and defect sites, which enhances NIR absorption and facilitates the diffusion of oxygen out of the material. We then have used time-resolved electron energy loss spectroscopy to measure that this augmentation results in fast and efficient oxygen removal from the material within 500 ns after irradiation with a NIR laser pulse. While localized non-thermal effects in the presence of electron-beam–induced defects cannot be entirely excluded, the collective experimental evidence indicates that the reduction is predominantly photothermal. We estimate that in the process, the sample achieves a temperature of 206 ºC for a duration of up to 2 milliseconds, which is enough to begin the formation of short-range graphitic order. The single-pulse NIR response of GO is therefore best described as a thermally activated process occurring within a beam-modified chemical environment, where transient laser heating drives oxygen desorption and structural reorganization rather than direct photochemical bond cleavage.

This work adds to a growing amount of evidence suggesting that excitation of GO by high-energy electrons and photons promotes the photoreduction process. This has been shown to be the case for the photoreduction of GO in an aqueous environment, as the process was found to be initiated by photogenerated solvated electrons [21]. Furthermore, others have shown that pre-irradiation of GO films with UV laser pulses improved the efficiency of photoreduction using near-IR pulses[25]. The dramatic enhancement of the efficiency presented here suggests that it is possible that the defect-mediated photoreduction mechanism is more universal, as pre-existing defect concentrations or defects created by irradiation with intense laser beams focused on the sample surface can influence the optical absorbance and removal of oxygen species. Careful consideration of these effects is important to rationalize the large variation in photoreduction parameters presented in the literature and uncover the true nature of the photoreduction process of GO.

Finally, while this work points to a route for rapid photoreduction of GO, several challenges must be resolved before it can become deployable on an industrial scale. Scaling up the process requires further investigations in raster scanning e-beam lithography systems that are modified to incorporate infrared laser light. Furthermore, the electron beam dose and peak laser power should be further explored to avoid film rupture and provide more heat to the system to develop long-range order. Further investigations of this kind are ongoing to explore the interplay of these parameters in producing high-quality rGO.

**Acknowledgement**

We thank Jesus Angel Valdez at the McGill University Facility for Electron Microscopy Research for help in microscope operation and data collection. This work is supported by the Natural Sciences and Engineering Research Council of Canada (RGPIN-2021-03797), Fonds de Recherche du Québec-Nature et Technologies (ERP-2024-NC-329200 and PBEEE-2023-334793), and Canada Foundation for Innovation and Ministère de l’Economie et de l'Innovation du Québec (Project 31018). The authors acknowledge financial support from the CNRS-CEA METSA French network (FR CNRS 3507) on the platform IPCMS-UTEM and the Interdisciplinary Thematic Institute ITI-QMAT at the University of Strasbourg.

## Declaration of competing interest

The authors declare the following financial interests/personal relationships which may be considered as potential competing interests: Kenneth R Beyerlein reports financial support was provided by Quebec Research Fund Nature and Technology. Kenneth R Beyerlein reports financial support was provided by Natural Sciences and Engineering Research Council of Canada. If there are other authors, they declare that they have no known competing financial interests or personal relationships that could have appeared to influence the work reported in this paper.

## Author contribution

I.A. : Methodology, Formal Analysis, Investigation, Writing- Original Draft, Visualization; H.M. : Formal Analysis, Investigation, Writing – Review & Editing; M.P. : Methodology, Validation, Investigation, Resources, Writing – Review & Editing, Project Administration; S.V. : Software, Formal Analysis, Writing – Review & Editing; F. B. : Conceptualization, Resources, Writing – Review & Editing, Supervision, Project Administration, Funding Acquisition; K.R.B: Conceptualization, Methodology, Investigation, Resources, Writing- Original Draft, Writing – Review & Editing, Supervision, Project Administration, Funding Acquisition.

## Supporting information

The Supporting Information contains supplemental figures of EELS measurements and TEM micrographs and is available free of charge at.

**Fast reduction of electron-beam-activated graphene oxide by an infrared laser pulse**

**TOC figure**

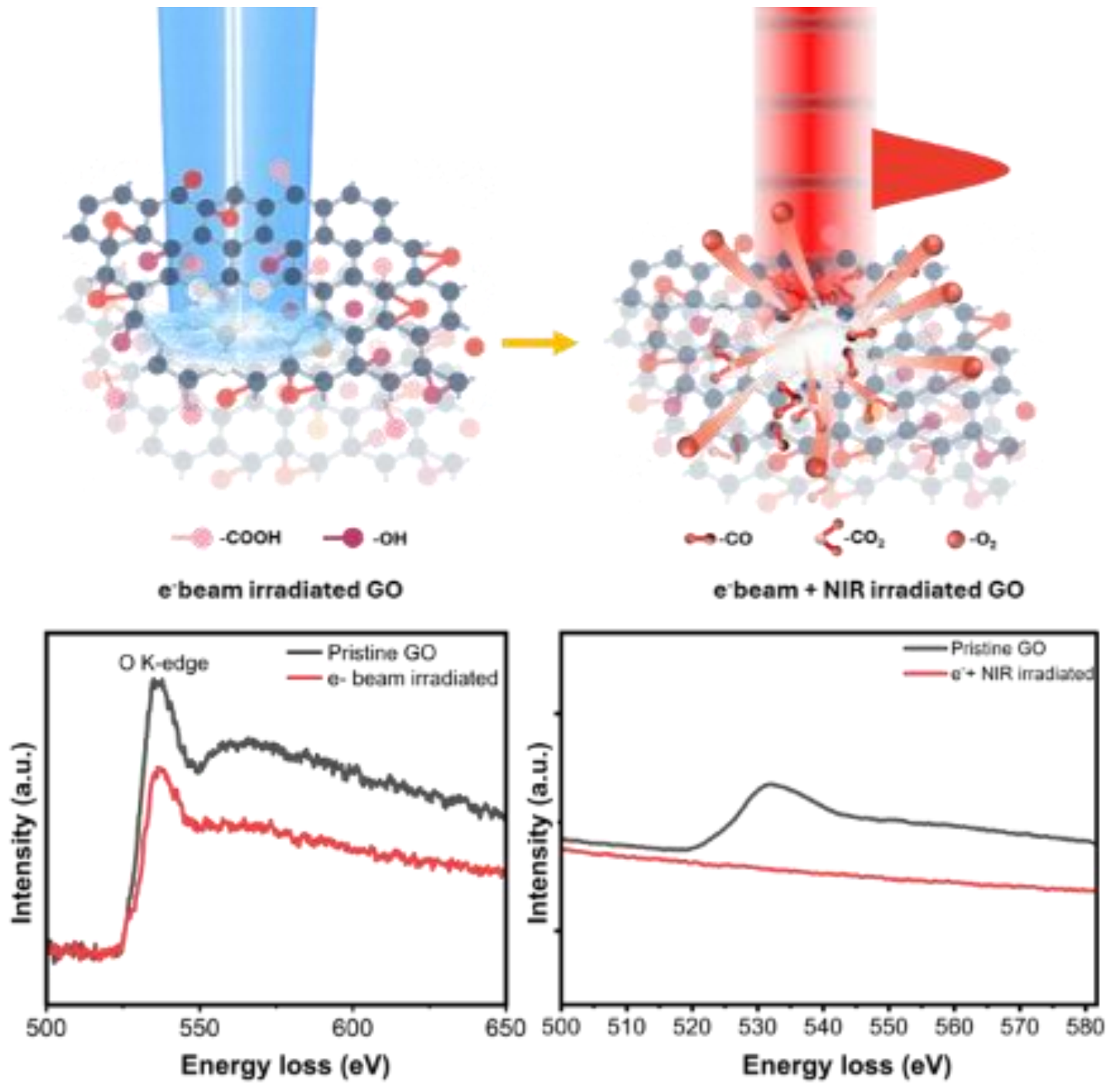

# Fast reduction of electron-beam-activated graphene oxide by an infrared laser pulse

Israt Ali [a], Hilaire Mba [b], Matthieu Picher [b], Shruti Verma [a], Florian Banhart [b], and Kenneth R. Beyerlein [a]*

[a] Institut National de la Recherche Scientifique (INRS), Center Énergie Matériaux Télécommunications, Varennes, Québec, J3X 1P7, Canada

[b] Institut de Physique et Chimie des Matériaux, UMR 7504, Université de Strasbourg, CNRS, 67034 Strasbourg, France

E-mail. Kenneth.beyerlein@inrs.ca

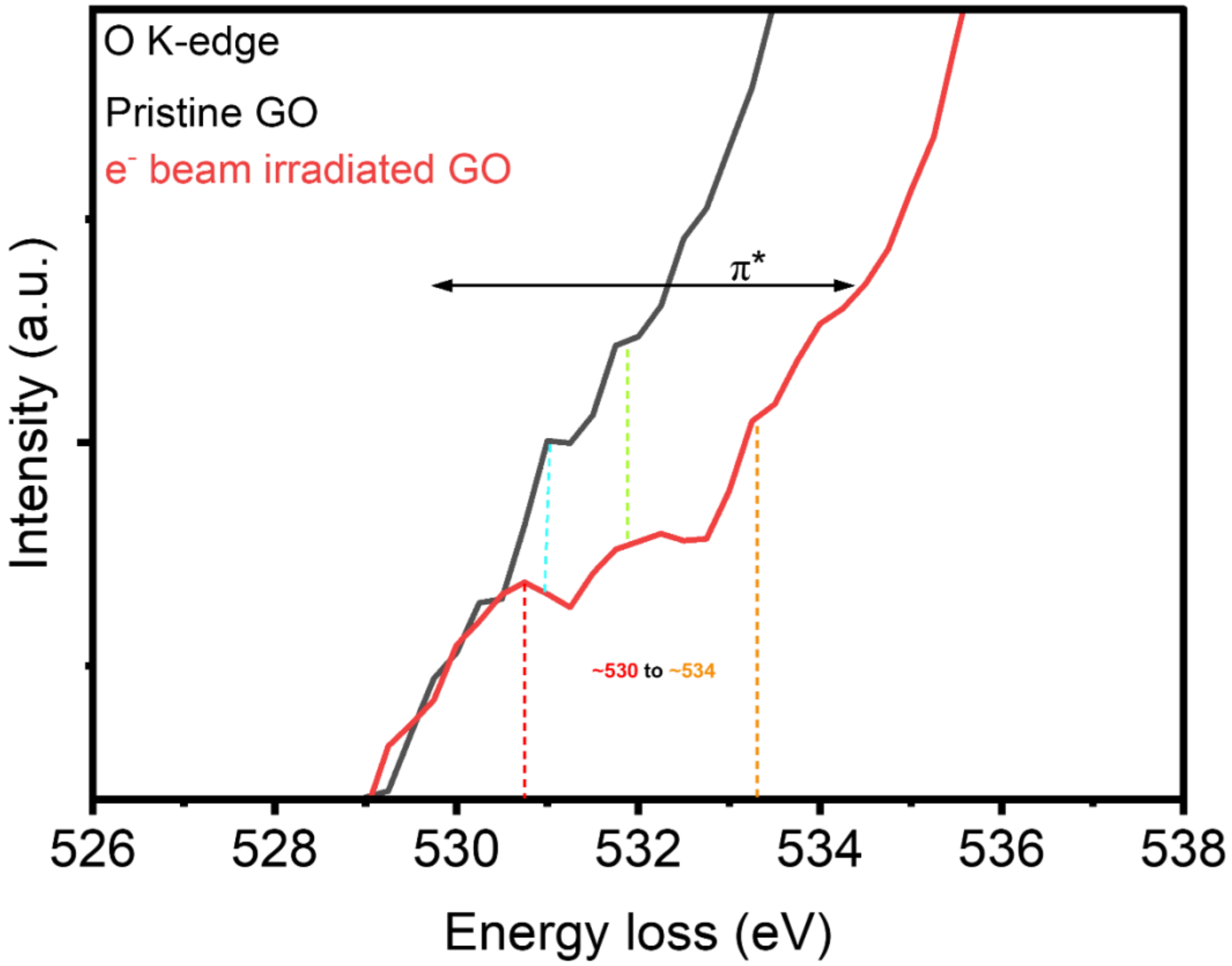

**Figure SI1.** Core-loss EELS spectra of the O K-edge before and after electron-beam irradiation. Core-loss electron energy-loss spectra acquired from graphene oxide (GO) before and after exposure to the electron beam show clear modifications in the O K-edge fine structure. The irradiated region exhibits a discernible shift in the edge onset together with changes in the relative intensities and shapes of the characteristic π* features, consistent with partial removal or rearrangement of oxygen-containing functional groups. These spectral variations confirm the beam-induced chemical modification of GO under the irradiation conditions used.

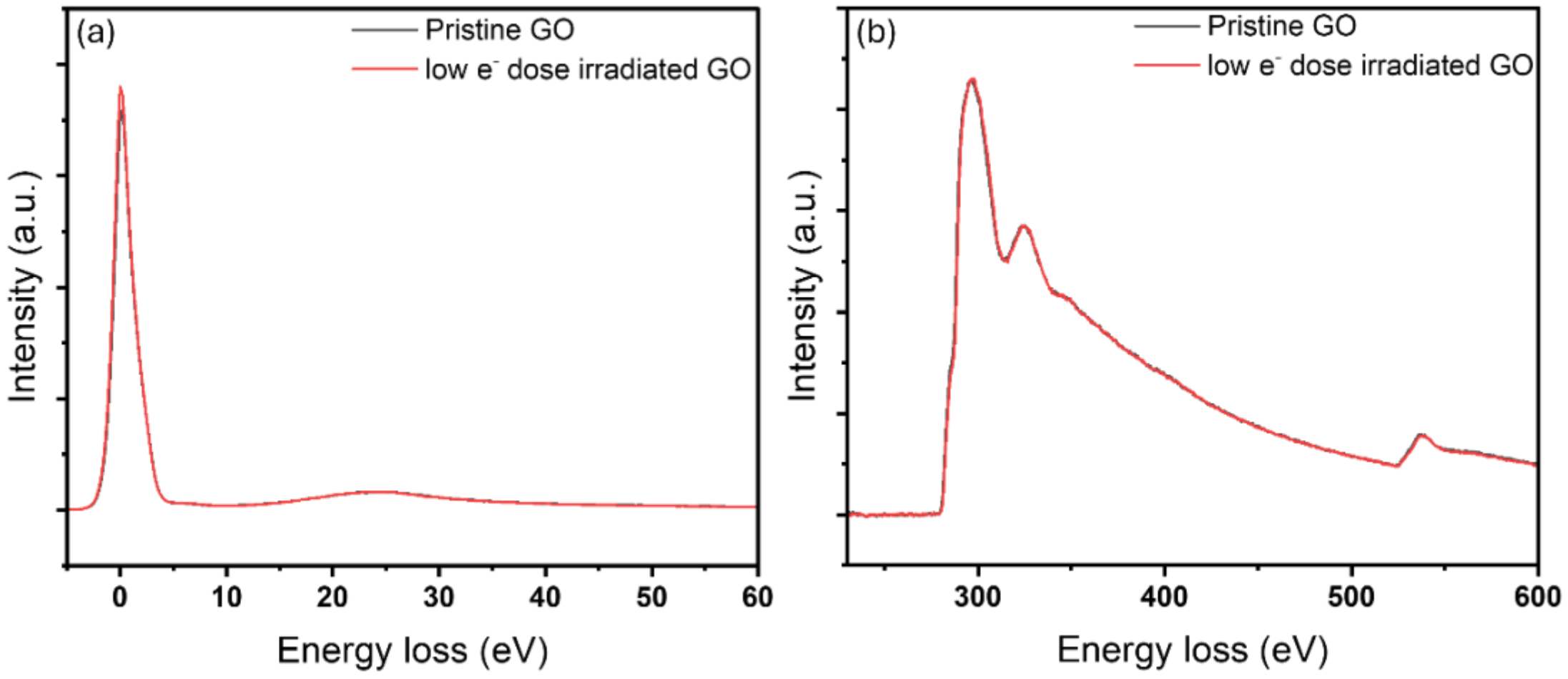

**Figure SI2**. EELS characterization of graphene oxide (GO) irradiated with a low electron dose ($9.35 \times 10^4$ $e^-/nm^2$) followed by a single NIR laser pulse (1064 nm, 7 ns, 181 $mJ/cm^2$). Note that while the electron dose used here is lower than the threshold found for sample activation ($6x10^7$ $e^-/nm^2$), the NIR laser fluence is nearly twice that used in other measurements. (a) Low-loss EELS spectra acquired before and after combined irradiation, showing no appreciable change in the plasmon response, indicative of the absence of significant structural modification, despite the higher laser fluence. (b) Core-loss EELS spectra showing the C K-edge and O K-edge regions before and after irradiation; the oxygen edge is clearly retained at full intensity in the post-irradiation spectrum, confirming that no measurable deoxygenation has occurred under these conditions.

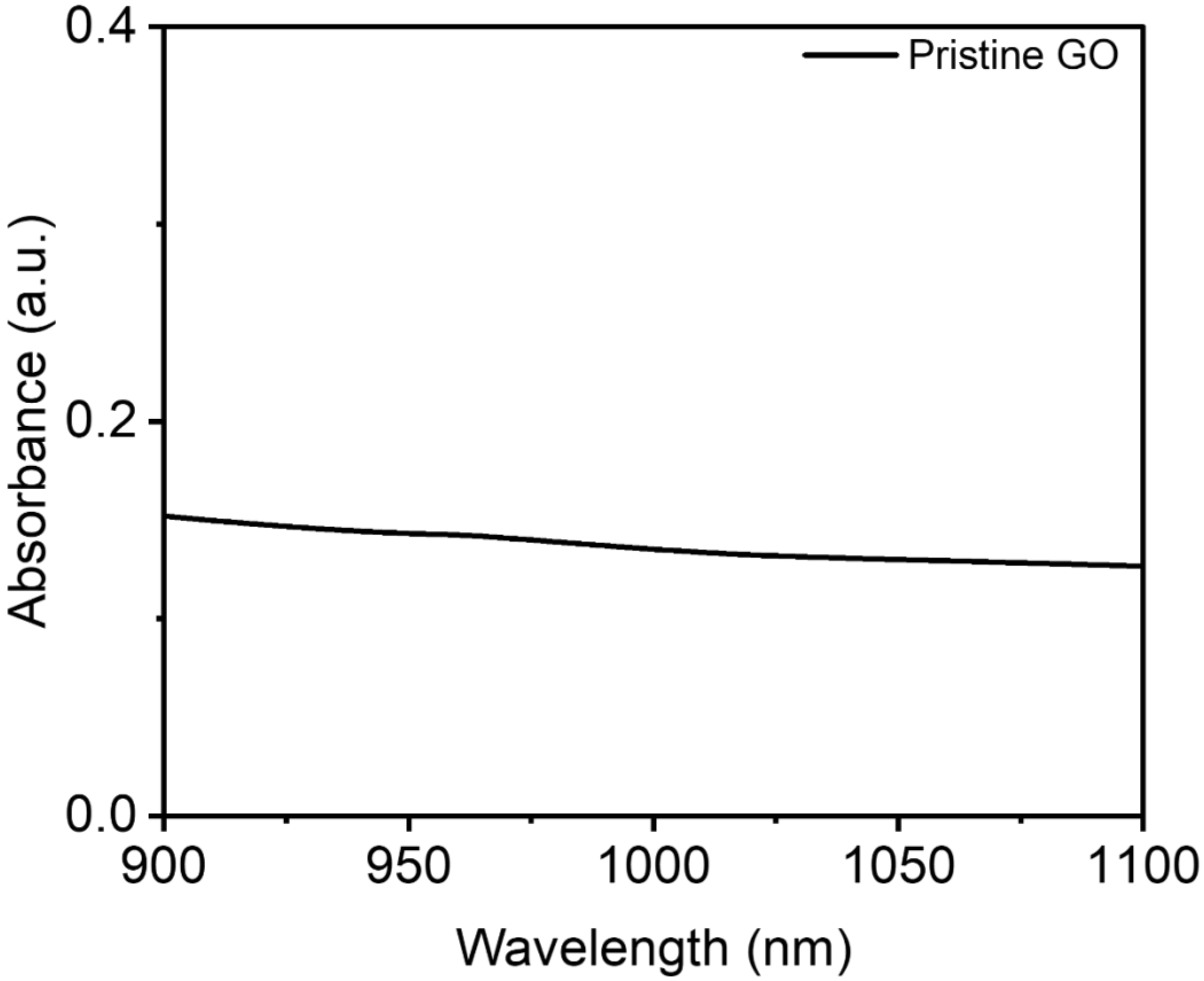

**Figure SI3.** The NIR absorption spectrum of a pristine GO film.

For pristine GO, the absorption coefficient was independently determined using UV–visible spectroscopy. The absorption coefficient, $\alpha$, was calculated from the measured absorbance, $A$, according to

$$\alpha = \frac{2.303A}{t}, \quad (1)$$

where t is the sample thickness which was 1 micrometer for this UV-Vis measurement.

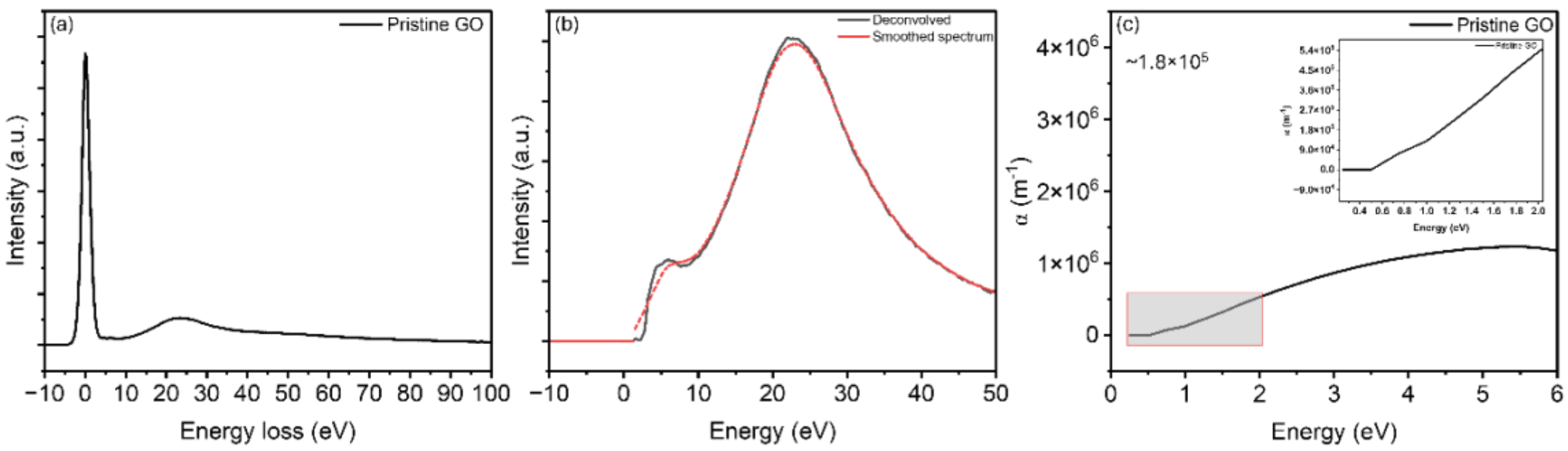

**Figure SI4**. Low-loss EELS analysis and absorption coefficient extraction for pristine GO. (a) Low-loss EELS spectrum acquired from GO (b) Corresponding deconvolved and smoothed low-loss spectrum used as input for the Kramers–Kronig analysis (KKA), showing the characteristic plasmon and interband transitions of the modified material. (c) Absorption coefficient spectrum obtained from KKA, plotted as a function of energy. The inset highlights the low-energy region, where the absorption coefficient reaches approximately $1.8\times10^5$ $m^{-1}$ near ~1.16 eV.

The absorption coefficient, $\alpha(E)$, was calculated from the extinction coefficient k(E) obtained via Kramers–Kronig analysis of the low-loss EELS spectra using the relation.

$$\alpha(E) = \frac{2E}{\hbar c}\, k(E) \quad (2)$$

where E is the photon energy, $\hbar$ is the reduced Planck constant, and c is the speed of light in vacuum. This expression links the energy-dependent optical absorption directly to the imaginary part of the complex refractive index and enables quantitative determination of the absorption coefficient from electron energy-loss measurements.

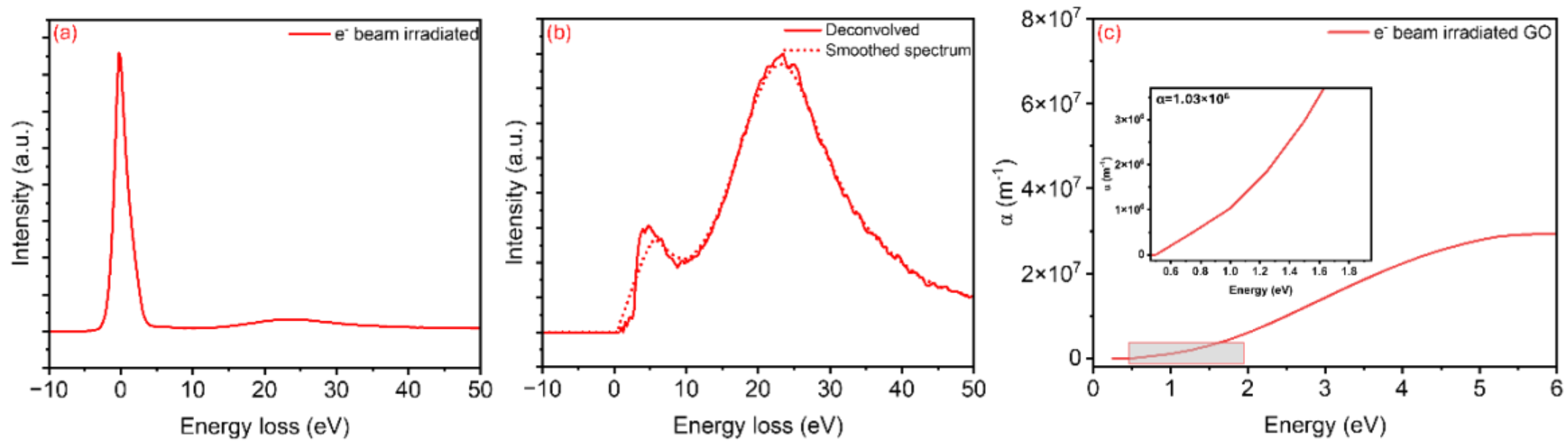

**Figure SI5.** Low-loss EELS analysis and absorption coefficient extraction for electron-beam-irradiated GO. (a) Low-loss EELS spectrum acquired from GO after electron-beam irradiation. (b) Corresponding deconvolved and smoothed low-loss spectrum used as input for the Kramers–Kronig analysis (KKA), showing the characteristic plasmon and interband transitions of the modified material. (c) Absorption coefficient spectrum obtained from KKA, plotted as a function of energy. The inset highlights the low-energy region, where the absorption coefficient reaches approximately $1\times10^{6}$ $m^{-1}$ near ~1.16 eV.

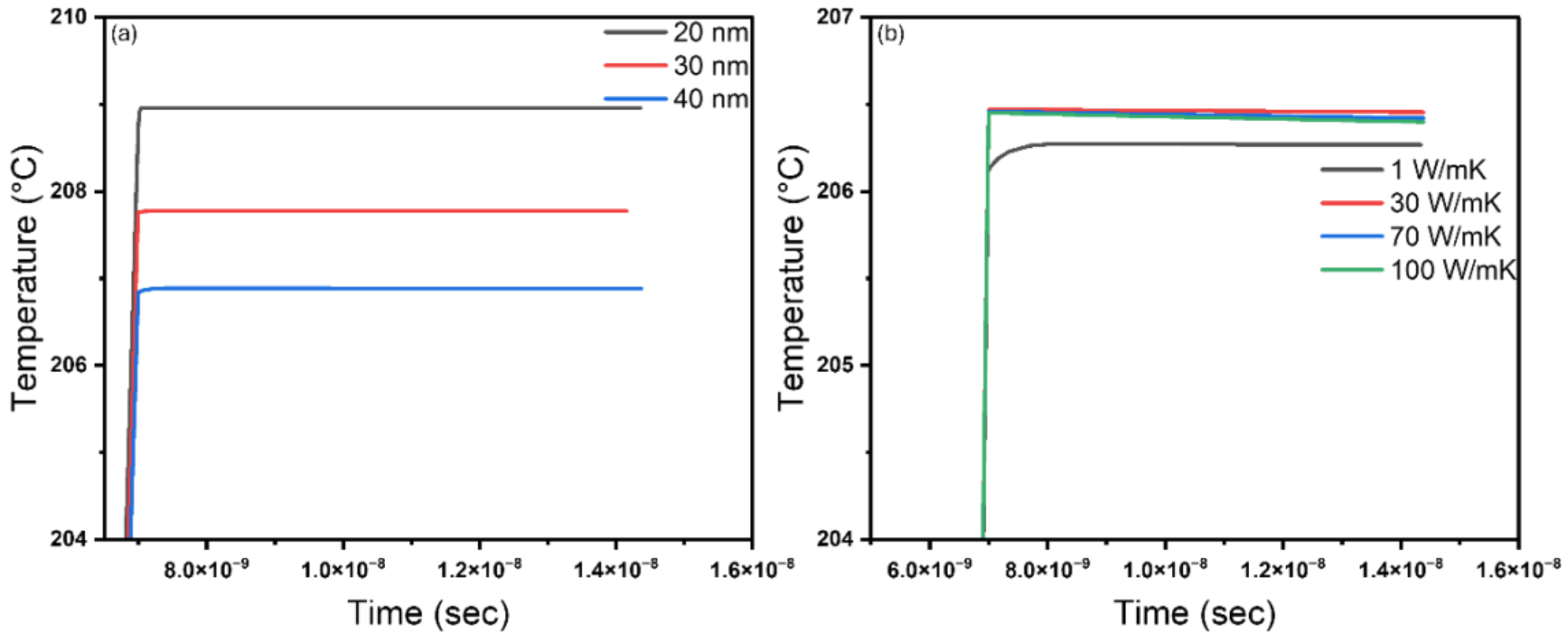

**Figure SI6**. COMSOL Multiphysics thermal simulation of laser-induced temperature rise in rGO films as a function of (a) film thickness (20, 30, and 40 nm) and (b) thermal conductivity (1, 30, 70, and 100 W/mK).

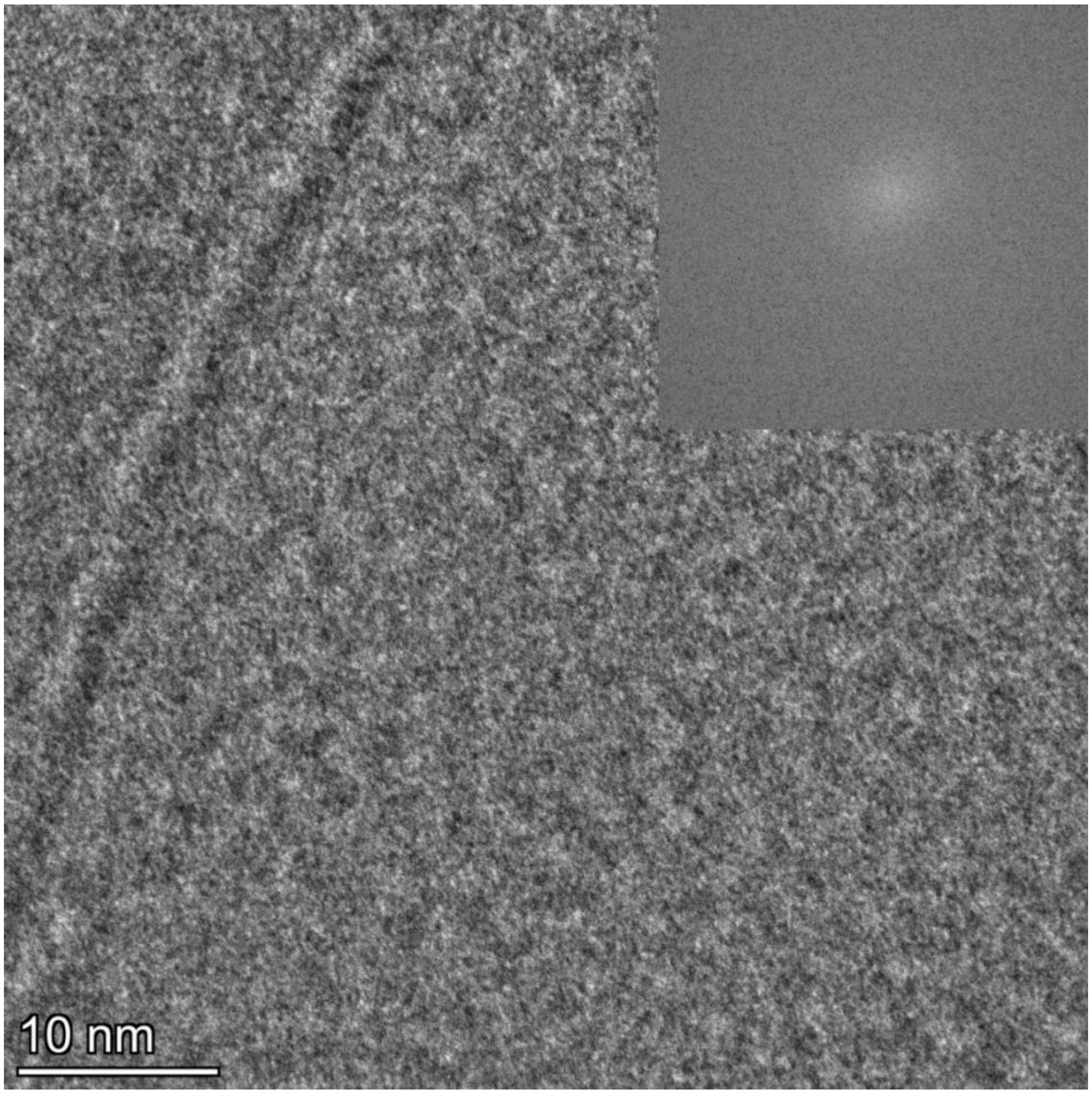

**Figure SI7.** HRTEM image of pristine graphene oxide, showing a predominantly amorphous structure with no discernible lattice fringes, with the inset showing the fast Fourier transform (FFT).

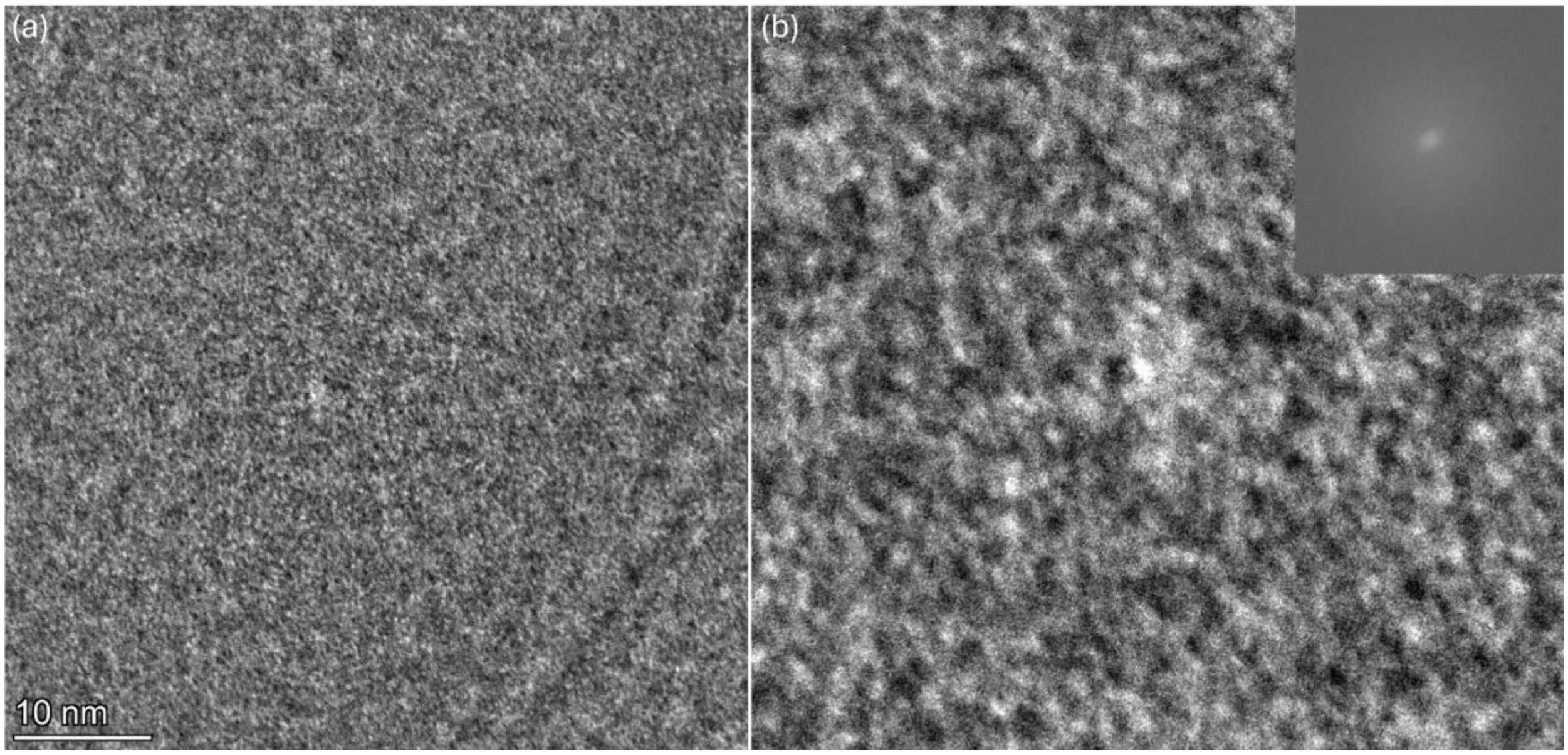

**Figure SI8.** HRTEM image of an electron-beam-irradiated GO region.

High-resolution TEM imaging of the electron-beam-irradiated area shows no discernible lattice fringes or periodic contrast indicative of crystalline ordering. The absence of long-range or short-range lattice coherence suggests that electron-beam exposure alone does not induce graphitization under the applied conditions. A localized bright contrast feature is observed, which may be evidence of a beam-induced local structural inhomogeneity, however its nature is unclear as its atomic structure is not evident.

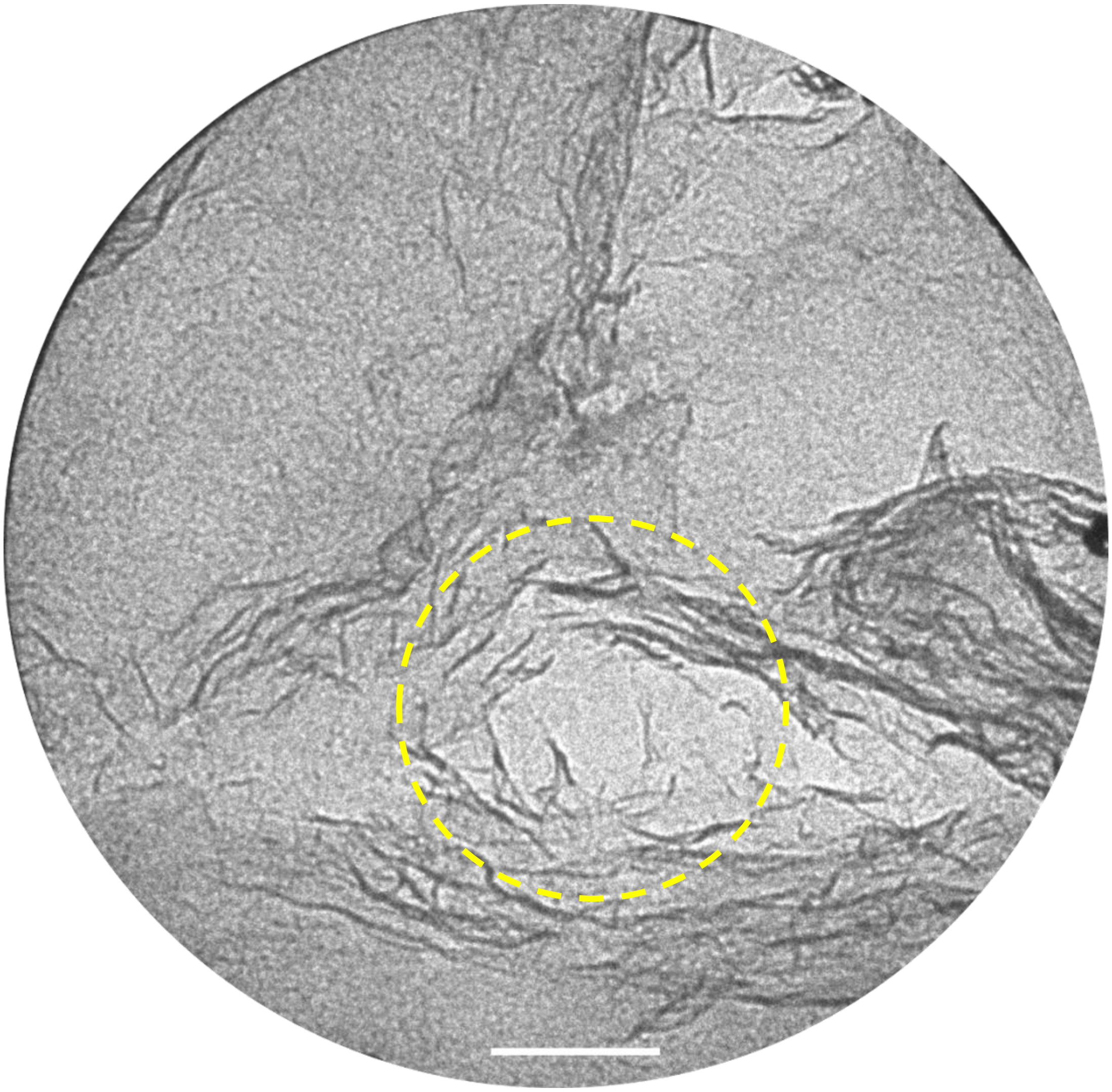

**Figure SI9.** TEM evidence of structural disruption induced by electron beam and single NIR pulse irradiation. TEM imaging of the irradiated region reveals pronounced surface rupture and material loss, consistent with rapid oxygen removal from within the GO film. The observed morphological damage in the area indicated by the dashed yellow circle suggests that beam-induced deoxygenation generates localized stress and outgassing, leading to mechanical failure and mass loss in the electron-irradiated area, followed by a single NIR pulse (scale bar 200 nm).

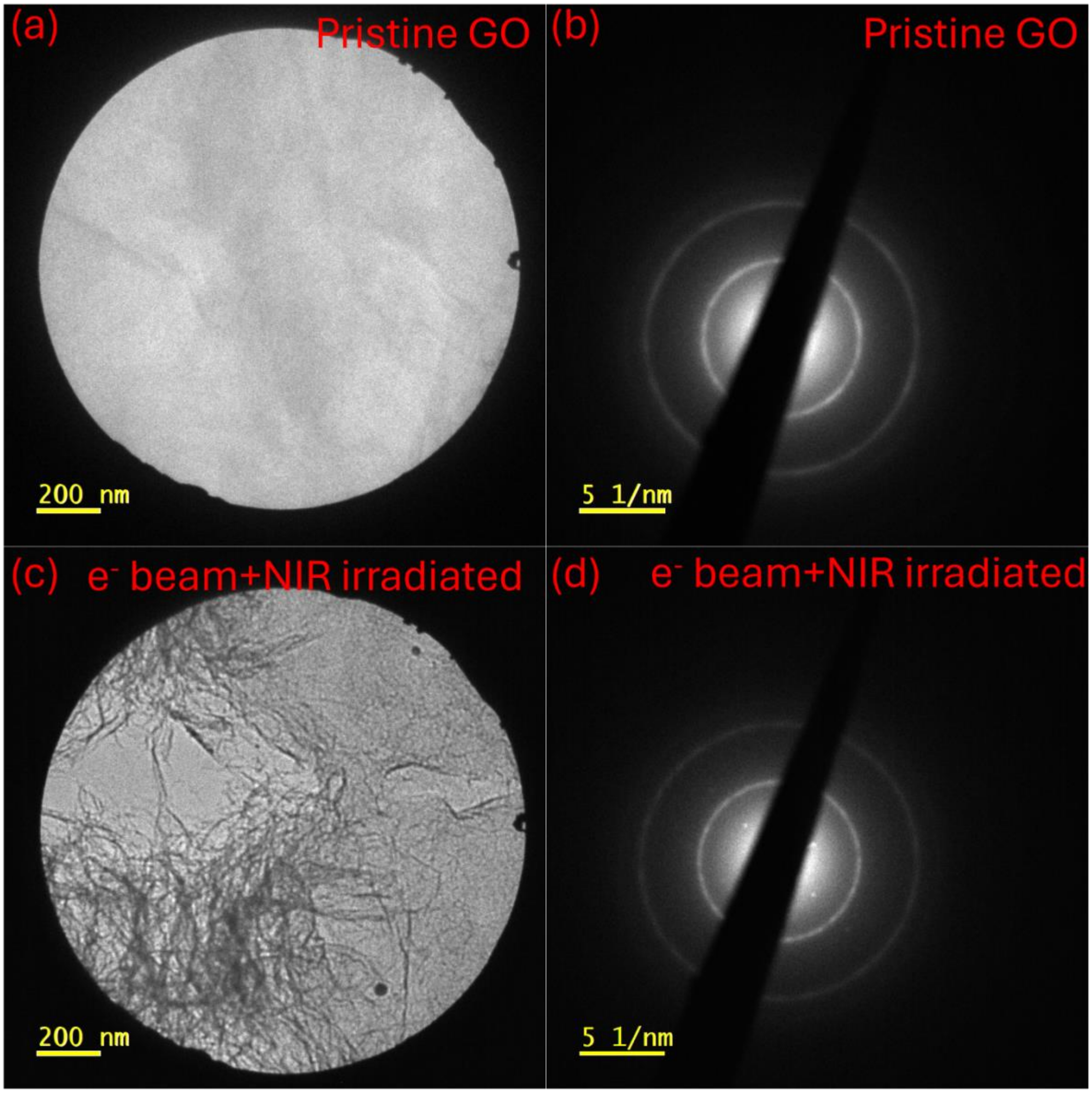

**Figure SI10**. Bright-field TEM images and corresponding SAED patterns of (a, b) pristine GO and (c, d) $e^-$ beam+NIR-irradiated rGO. (a, c) Low-magnification TEM images showing the selected-area regions used for diffraction analysis; scale bar: 200 nm. (b, d) Corresponding SAED patterns extended to include the (002) reflection region. The pristine GO pattern (b) shows a broad, diffuse ring, consistent with GO's disordered, turbostratic structure. After $e^-$ beam+NIR irradiation (d), the corresponding ring becomes sharper with a discernible spotty character, indicating the emergence of local (002) graphitic stacking order, with a measured d-spacing of ~0.353 nm obtained from the radial intensity profile.

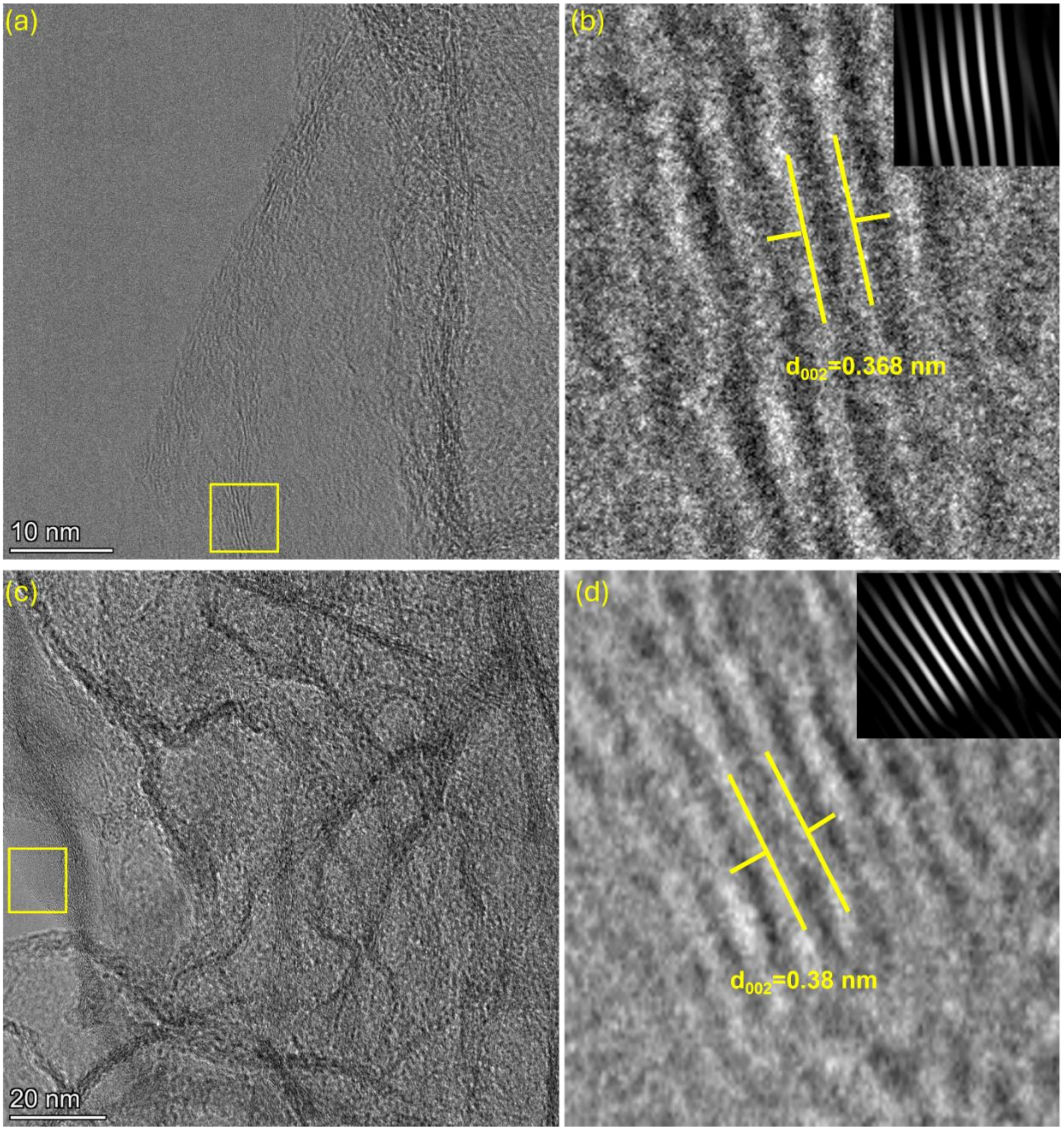

**Figure SI11**. HRTEM images of rGO sheets showing lattice fringes measured from unambiguous sheet-like morphologies. (a) Low-magnification HRTEM image showing a folded rGO sheet edge, with the boxed region magnified in (b). (b) Magnified view showing parallel lattice fringes with a measured interplanar spacing of $d_{002}$ = 0.368 nm; inset shows the corresponding FFT pattern displaying a single periodic set of spots consistent with graphitic layer stacking. (c) Low-magnification HRTEM image of another folded rGO sheet region, with the boxed region magnified in (d). (d) Magnified view showing lattice fringes with a measured interplanar spacing of $d_{002}$ = 0.38 nm; inset shows the corresponding FFT pattern. The slightly expanded $d_{002}$ spacing relative to pristine graphite (0.335 nm) is attributed to residual oxygen-containing functional groups and structural disorder in rGO.